\documentclass[%
 reprint,
 amsmath,amssymb,
 aps,
]{revtex4-2}

\usepackage{graphicx}% Include figure files
\usepackage{dcolumn}% Align table columns on decimal point
\usepackage{bm}% bold math
\usepackage{xcolor}
\usepackage{comment}
\usepackage{amsmath}
\definecolor{darkgreen}{rgb}{0,0.6,0}

\begin{document}

\preprint{APS/123-QED}

\title{Space-resolved dynamic light scattering within a millimetric drop: from Brownian diffusion to the swelling of hydrogel beads}% Force line breaks with \\

\author{Matteo Milani$^1$}
% \altaffiliation[Also at ]{}%Lines break automatically or can be forced with \\
\author{Ty Phou$^1$}%
\author{Guillame Prevot$^1$}
%\author{Christian Ligoure}%
%\author{Chiara Kindelberger}
\author{Laurence Ramos$^1$}%
\email{laurence.ramos@umontpellier.fr}
\author{Luca Cipelletti$^{1,2}$}%
\email{luca.cipelletti@umontpellier.fr}

\affiliation{%
$^1$Laboratoire Charles Coulomb (L2C), Universit\'e Montpellier, CNRS, Montpellier, France\\
$^2$Institut Universitaire de France, Paris, France
}%


%\affiliation{
%\affiliation{
%\affiliation{%


\date{\today}% It is always \today, today,

\begin{abstract}
We present a novel dynamic light scattering setup to probe, with time and space resolution, the microscopic dynamics of soft matter systems confined within millimeter-sized spherical drops. By using an ad-hoc optical layout, we tackle the challenges raised by refraction effects due to the unconventional shape of the samples. We first validate the setup by investigating the dynamics
of a suspension of Brownian particles. The dynamics measured at different positions in the drop, and hence different scattering angles, are found to be in excellent agreement with those obtained for the same sample in a conventional light scattering setup. We then demonstrate the setup capabilities by investigating a bead made of a polymer hydrogel undergoing swelling. The gel microscopic dynamics exhibit a space dependence that strongly varies with time elapsed since the beginning of swelling. Initially, the dynamics in the periphery of the bead are much faster than in the core, indicative of non-uniform swelling. As the swelling proceeds, the dynamics slow down and become more spatially homogeneous. By comparing the experimental results to numerical and analytical calculations for the dynamics of a homogeneous, purely elastic sphere undergoing swelling, we establish that the mean square displacement of the gel strands deviates from the affine motion inferred from the macroscopic deformation, evolving from fast diffusive-like dynamics at the onset of swelling to slower, yet supradiffusive, rearrangements at later stages.  
\end{abstract}

\maketitle

%\tableofcontents

\section{\label{sec:level1}Introduction}

Soft matter systems comprising colloidal particles, emulsion drops, foam bubbles and polymer and surfactant molecules are ubiquitous in industrial processes and products and are actively studied in fundamental research~\cite{barrat_soft_2024}. The interest in soft systems owes to their high susceptibility to even modest external stimuli, and to their typical structural length scale comparable to the wavelength of visible light, which imparts them interesting optical properties and makes them suitable for powerful experimental techniques  such as optical microscopy and light scattering. Furthermore, the tunability of the interactions between the building blocks of soft materials allows for a large variety of structures, resulting in a wide spectrum of physical, mechanical, and optical properties.

Research in soft matter initially focused mostly on systems at thermodynamic equilibrium, while in the last decades the investigation of out-of-equilibrium systems, including active matter, has tremendously gained momentum~\cite{noauthor_out--equilibrium_2023,bischofberger_editorial_2023}. 

Soft matter, both at and out of equilibrium, is typically investigated using bulk samples: the behavior of soft systems confined or manufactured in specific shapes, such as in thin films or spherical drops, has received comparatively less attention. However, these systems typically have distinctive features as compared to their bulk counterparts, due to the larger surface-to-volume ratio and specific boundary conditions, leading to fascinating behavior and remarkable structural and use properties. 

One prototypical example is provided by drops of colloidal suspensions, whose drying has been investigated in fundamental studies ~\cite{tsapis2005onset,basu_towards_2016,boulogne_buckling_2013,bamboriya2023universality,milani2023double}, in view of its interest in biology, e.g. in virus survival ~\cite{huynh2022evidence} or blood desiccation~\cite{brutin2011pattern}, as well as for applications~\cite{sadek_drying_2015}, including the design of functional supraparticles~\cite{liu2019tuning,wooh_synthesis_2015}. Another example of popular spherical soft systems are hydrogel beads. They are widespread in applications ranging from food science~\cite{saqib2022hydrogel,patel2022hydrogels} to the adsorption of dyes~\cite{chatterjee2010adsorption}, acids~\cite{yan2005adsorption} or metals~\cite{li2005copper}, to drug delivery~\cite{jain2007design}. 

The behavior of drop- or bead-like samples is often intimately related to their very shape and to the absence of interactions with a sample container, in contrast to conventional measurements on bulk materials. For instance, polymeric hydrogels exhibit a wide variety of behaviors such as shrinkage upon gelation,~\cite{lin2020effect}, slow syneresis after gelation~\cite{velings1995physico}, swelling when immersed in a good solvent~\cite{tanaka1979kinetics,patel1989mechanical,hong2008theory,barros2012surface,bertrand2016dynamics}, which are different in a bulk sample and in beads. Similarly, the interactions with a substrate were shown to play a key role in the evolution of drying drops of colloidal suspensions~\cite{kim2015crack,basu_towards_2016}.  

While there is a clear interest in characterizing the structure and dynamics of spherically confined soft systems, the available experimental tools are far more limited than for bulk samples. Methods used in previous works include low magnification video imaging and shadowgraphy~\cite{tsapis2005onset,bansal_universal_2015,bertrand2016dynamics}, imaging of a portion of the drop by confocal microscopy~\cite{wooh_synthesis_2015,sperling_understanding_2016}, post-mortem electron microscopy~\cite{tsapis2005onset,miglani_sphere_2015,bansal_universal_2015}, synchrotron X-ray scattering~\cite{sen_slow_2007,sen_probing_2014}, and nuclear magnetic resonance (NMR)~\cite{engelsberg2013free,barros2012surface}.
These methods provide valuable information, however they are restricted to structural quantities, they sometimes lack spatial resolution (e.g. for x-ray scattering) and may be restricted to some portions of the drop (e.g. in confocal micorscopy) or give no access to the system evolution (as in post-mortem electron microscopy).

Measurements of the dynamics appear as an appealing alternative or complement to structural studies. Dynamic light scattering (DLS)~\cite{berne2000dynamic} is a well-established and popular method for measuring the microscopic dynamics of soft systems. However, in conventional DLS intensity correlation functions, which quantify the microscopic dynamics, are obtained by extensively averaging over time a signal issued from the whole sample volume illuminated by the incident laser beam. In other words, conventional DLS lacks both space and time resolution, two key features for investigating spherically confined systems, where boundary conditions (and hence distance to the surface) play a key role, and where time-dependent phenomena, such as evaporation or swelling are ubiquitous.

Photon Correlation Imaging (PCI)~\cite{duri2009resolving} is a variant of conventional DLS that affords both spatial and temporal resolution. It has been used to characterize the dynamics of bulk samples, both spontaneous~\cite{duri2009resolving,maccarrone_ultra-long_2010,buzzaccaro_spatially_2015,lattuada_spatially_2021} and under a mechanical drive~\cite{crassous_diffusive_2009,pommella2019coupling,aime_unified_2023}. Very recently, we have reported a study of the drying of drops of colloidal suspensions based on PCI~\cite{milani2023double}. Here, we present a detailed description, validation and demonstration of the PCI setup used therein. 

The rest of the paper is organized as follows: in Sec.~\ref{Materialsandmethods} we introduce the PCI setup, discuss how to account for refraction effects associated with the spherical shape of the sample, define the quantities introduced to characterize the microscopic dynamics, and present measurements of the size of the scattering volume and the setup stability. In Sec.~\ref{sec:Brownian} we validate the setup and the data analysis procedure by measuring diluted colloids undergoing Brownian motion. Section~\ref{sec:polymergel} deals with the microscopic dynamics of a polymer gel upon swelling. We first discuss theoretically the behavior expected for uniform, affine swelling, a benchmark against which experimental data are then compared. Our findings for the polymer gel are compared to previous works in Sec.~\ref{sec:discussion_polymer}. Finally, Sec.~\ref{sec:conclusions} provides some concluding remarks. Details on some calculations and data analysis methods are discussed in Appendixes~\ref{app:qvector}-\ref{app:Igel}.

\section{Photon Correlation Imaging setup}\label{Materialsandmethods}
\subsection{Optical layout}\label{sec:optical_layout}

%%%%%%%%%%%% FIG 1 %%%%%%%%%%%%%%%%%%%%%%%
\begin{figure*}
\includegraphics[width=1\linewidth]{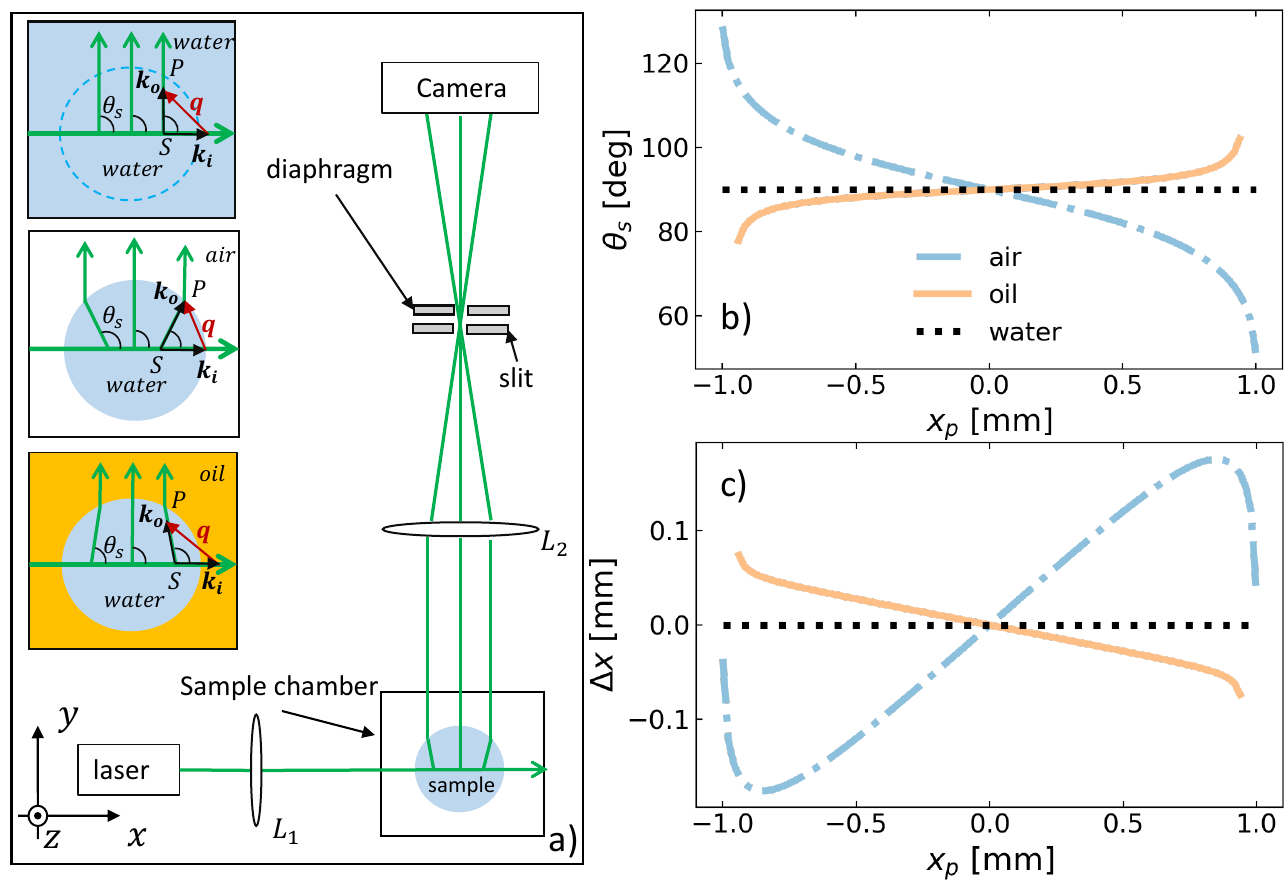}% Here is how to import EPS art
\caption{\label{fig:set-up} (a) Scheme of the dynamic light scattering setup optimized for spherical, millimiter-sized samples. The incoming laser beam propagates along the $x$ axis and is mildly focused by the lens $L_1$ in the middle of the sample. The lens $L_2$ forms an image of the scattering volume on the sensor of a CMOS camera. By combining a vertical slit, parallel to the $z$ direction, and a diaphragm, we ensure that the image is formed by scattered light exiting the drop along the $y$ direction, to within acceptance angles of $0.55$ deg and  $1.65$ deg in the $(x,y)$ and $(y,z)$ planes, respectively. The sample is placed in a cubic glass box of side $5$ cm. Insets: schematic top view of the relevant optical rays for a water-based spherical sample surrounded by water, air and oil, respectively. Due to refraction, the scattering angle $\theta_s$ for light exiting the drop along the $y$ direction depends on the $x$ coordinate of the scatterer. (b) Scattering angle $\theta_s$ as a function of $x_p$, the $x$ coordinate of the point $P$ on the bead surface shown in the insets of a). Data for a bead of radius $1$ mm surrounded by air, oil or water, as shown by the labels. (c) Difference $\Delta x = x_P - x_S$  as a function of $x_P$, where $x_S$ is the abscissa of the point $S$ from which a scattering ray originates, and $x_P$ is the abscissa of the point $P$ where the scattered ray intersects the surface of the sphere. The curves refer to the case of a bead of $1$ mm radius surrounded by air, oil or water, see legend in b).}
\end{figure*}
%%%%%%%%%%%%%%%%%%%%%%%%%%%%%%%%%%%%%%%%%%%%

Figure~\ref{fig:set-up}a shows schematically the light scattering setup, specifically designed to probe with space and time-resolution the microscopic dynamics within millimeter-sized spherical samples. Due to the small radius of curvature of the sample, whenever the refractive index of the solvent in the drop, $n$, is different from that of the surrounding medium, refraction effects considerably complicate the definition of the scattering angle. We minimize these complications by illuminating the sample with a thin beam impinging perpendicularly to the drop surface, such that the incoming beam is essentially not refracted, and by collecting light exiting the sample along a single, well-defined direction.  

The light source is a single-mode continuous wave laser (Verdi V-2 by Coherent) operating at an in-vacuo wavelength $\lambda_0 =532.5$ nm with a maximum power of $2$ W. A first lens ($L_1$ in Fig.~\ref{fig:set-up}a, with diameter $65$ mm and focal length $64$ mm), collimates the beam down to a beam waist $w_0 = 59.5 \pm 0.1~\mu$m, much smaller than the typical drop radius (see Sec.~\ref{sec:setupcharacterization} for the measurement of $w_0$). Thus, the scattering volume is a thin cylinder centered around a drop diameter, along the $x$ axis direction. To achieve spatial resolution, the scattered light is not collected in the far field, as in conventional scattering, but rather by forming an image of the scattering volume on the detector of a CMOS camera, using lens L2 (diameter $40$ mm, focal length $82$ mm). To insure that only light propagating along the direction of the $y$ axis is collected by the camera, we place a slit and a diaphragm in the back focal plane of L2, which reject scattered light propagating at angles larger than $\theta_{yz} = 1.65$ deg and $\theta_{yx} = 0.55$ deg with respect to the $y$ axis, in the $(y,z)$ and $(x,y)$ planes, respectively. Note that the drop and the surrounding medium are contained in a cubic sample chamber with walls perpendicular to the $x$ and $y$ directions, so as to avoid refraction effects at the chamber-air interfaces. Finally, by taking the image of a ruler placed at the sample position, we determine the system magnification to be $M = 2.11 \pm 0.02$. The combination of the illumination and collection optical layout described above allows for associating a well defined scattering angle $\theta_s$, and thus a well defined magnitude of the scattering vector $q = 4 \pi n \lambda_0^{-1} \sin(\theta_s/2$), to each point within the scattering volume, as we will detail in Sec~\ref{sec:RefractionEffects}.

Depending on the sample dynamics, we use either a fast camera (Phantom Miro M310 by AMETEK Vision Research, e.g. for the Brownian dynamics described in Sec.~\ref{sec:Brownian}) or a conventional camera (acA2000 340km by Basler, e.g. for the polymer gel of Sec.~\ref{sec:polymergel}). The fast camera is run at $19000$ fps, for a typical run duration of 60 ms. The size of the acquired images is $896$ pixels $\times 184$ pixels, and the pixel size is $20 \times 20 ~\mu\mathrm{m}^2$. By contrast, to measure slow dynamics we use a CMOS camera (acA2000 340km by Basler). In this case the collected images have a full frame size of $2048$ pixels $\times 1088$ pixels and a pixel size of $5.5$ $\mu$m. To allow for faster image acquisition, we typically acquire only a fraction of the full frame, selecting a region of interest of $2048$ pixels $\times 512$ pixels. Here the acquisition is taken using the variable delay scheme of Ref.~\cite{philippe2016efficient}, with shortest delay 1 s.

\subsection{Refraction effects}\label{sec:RefractionEffects}

Due to its spherical shape, the sample itself acts as an optical component of the setup. As shown in the insets of Fig.~\ref{fig:set-up}a, the scattered light propagating towards the detector is refracted at the interface between the sample and the external medium, whenever the index of refraction of the surrounding medium is different from that of the sample. As a consequence, each ray emerging from the sample and directed towards the detector is associated with a particular scattering angle $\theta_{s}$, such that there is an intrinsic coupling between position and scattering vector $\textbf{q} = \textbf{k}_o - \textbf{k}_i$, where $\textbf{k}_i$ is the wave vector of the incident beam and $\textbf{k}_o$ that of the scattered light.

We now determine the components and magnitude of $\textbf{q}$ as a function of the coordinates $(x_P,y_P,z_P$) of a point $P$ at the intersection between a scattered ray collected by the camera and the drop surface, where we take the drop center as the origin of a Cartesian reference frame. The incident wave vector is $\textbf{k}_i = 2 \pi n \lambda_0 \hat{e}_x $, with $ \hat{e}_x $ the unit vector along the x axis. Thanks to the combination of the slit and diaphragm, only scattered light propagating along $\hat{e}_y$ after leaving the drop contributes to the image formed on the camera detector. The propagation direction of the scattered light within the drop, $\hat{k}_o=\textbf{k}_o/k_o$, is found by imposing Snell's law at the drop-surrounding medium interface. By introducing the unit vector pointing at $P$, $\hat{r}_P = (x_P\hat{e}_x +y_P\hat{e}_y+z_P\hat{e}_z)/\sqrt{x_P^2+y_P^2+z_P^2}$, after some tedious but straightforward calculations, one finds
\begin{equation} 
\label{eq:ko}
\hat{k}_o = a \hat{r}_P + b \hat{e}_y \,,
\end{equation}
with $a$ and $b$ two prefactors that depend on the position of $P$ and the refractive index mismatch between the drop and the surrounding medium, and whose expression is given in Eqs.~\ref{eq:a_b} of Appendix~\ref{app:qvector}. Once $\textbf{k}_o$ is determined,
the scattering vector and the scattering angle are calculated according to
\begin{equation} 
\label{eq:qvector}
\textbf{q} = \textbf{k}_o-\textbf{k}_i\,,
\end{equation}
\begin{equation} 
\label{eq:scangle}
\theta_{s} =\arccos (\hat{k}_o\cdot\hat{e}_x)  \,,
\end{equation}
where we have used $\hat{k}_i = \hat{e}_x$.
Figure~\ref{fig:set-up}b shows the scattering angle $\theta_{s}$ as a function of $x_P$, for a water-based sample ($n=1.33$) in a one-millimeter drop surrounded by air ($n_{air}=1$), or by the oil used for the experiments of Sec.~\ref{sec:Brownian}, $n_{oil}=1.403$. Note that the trend is opposite for air and oil as a surrounding medium, and that the deviations with respect to $\theta_s = 90$ deg, the scattering angle for a ray originating from the center of the drop, can be as large as $30$ deg for a water-in-air drop. The corresponding scattering vectors have magnitude $14\mu$m$^{-1} \le q \le 28\mu$m$^{-1}$, with $q\approx22\mu$m$^{-1}$ for $\theta_s = 90$ deg. Since DLS techniques are sensitive to relative displacements of the order of $1/q$, our setup is sensitive to motion on length scales of the order of a few tens of nanometers. 

Another consequence of refraction at the drop surface is the fact that the coordinate $x_P$ where a scattered ray intersects the drop surface differs from the coordinate $x_S$ along the scattering volume from which that scattered light originates, see also Fig.~\ref{fig:scattering_for_theory}b. Note that $x_P$ is directly obtained from the position on the image collected by the CMOS camera, but the desired, physically relevant quantity is the position in the sample, i.e. $x_S$. Assuming for simplicity that the incident beam is infinitely thin, i.e. that the scattering originates from a point $S$ on the $x$ axis, and using $\textbf{OS} = \textbf{OP} - \textbf{SP}$, one finds
\begin{equation} 
\label{eq:x}
x_S = x_P +  \frac{\hat{k}_{ox}}{|1-\hat{k}_{ox}|}\ \sqrt{y_P^2 + z_P^2} \,,
\end{equation}
with $\hat{k}_{ox} = \hat{k}_{o} \cdot\ \hat{e}_x$. Figure~\ref{fig:set-up}c shows $\Delta x = |x_S-x_P|$, as a function of $x_P$: deviations as large as 15\% may be observed for a water drop in air. In the following, we systematically use the coordinate $x_S$ when showing space-dependent data.

%Note that, since DLS probes the relative motion of the scatterers projected on $\textbf{q}$ and given the particular geometry of the setup, it is important to consider the full vectorial expression of the scattering vector $\textbf{q}$ and not only its the modulus, $q$.

\subsection{Quantifying the dynamics: multispeckle intensity correlation functions}

To quantify the sample dynamics we take a time series of images of the speckle pattern imaged onto the CMOS detector. Each image is divided into several Regions of Interest (ROIs), typically rectangles of dimensions  $(30 \times100)$ pixels$^2$, corresponding to about $(70 \times 260)$ $\mu$m$^2$ in the sample. The size of the ROIs is chosen so that each ROI contains around $100$ speckles, which provides an acceptable statistical noise~\cite{maccarrone2010ultra}, while keeping the ROI size small enough to obtain about 10 ROIs over the bead diameter. The local dynamics within a given ROI is quantified by a two-time degree of correlation~\cite{cipelletti2002time}:
%%%%%%%%%%%%%%%%  EQ %%%%%%%%%%%%%%%%%%%%
\begin{equation}\label{eq:cI}
    c_{I}(t,\tau,x) = \alpha \frac{\langle I_p(t)I_p(t+\tau) \rangle_{x} }{\langle I_p(t)\rangle_{x}\langle I_p(t+\tau) \rangle_{x}}-1
\end{equation}
%%%%%%%%%%%%%%%%%%%%%%%%%%%%%%%%%%%%%%%%
with $x$ the position along the diameter of the sample, $\tau$ the delay between images, $I_p(t)$ the intensity of the $p$-th pixel of the ROI at time $t$, and $\langle  \cdot \cdot \cdot \rangle_{x}$ an average over all pixels within a ROI centred around $x$. The multiplicative factor $\alpha$ is chosen such that the intensity correlation function $g_2(\tau, x) - 1$, obtained by averaging $c_{I}(t,\tau,x)$ over a suitable time interval, tends to $1$  for $\tau\to 0$. 

For a dilute suspension of identical spheres of radius $a$ undergoing Brownian motion, the intensity correlation function is a simple exponential decay:
%%%%%%%%%%%%%%%%  EQ %%%%%%%%%%%%%%%%%%%%
\begin{equation}
    g_2(\tau)-1 =  \exp(-2q^2D_0\tau)= \exp\left(-2\frac{\tau}{\tau_E}\right)
\label{eq:g2brownian}
\end{equation}
%%%%%%%%%%%%%%%%%%%%%%%%%%%%%%%%%%%%%%%%%
where $\tau_E = (q^2D_0)^{-1}$, with $D_0=\frac{k_B T}{6\pi \eta a}$ the diffusion coefficient given by the Stokes-Einstein relation, with $k_B$ Boltzmann's constant, $T$ the temperature, and $\eta$ the solvent viscosity. More generally, $g_2-1$ may be conveniently expressed by a stretched or compressed exponential:
%%%%%%%%%%%%%%%%  EQ %%%%%%%%%%%%%%%%%%%% ~\ref{eq:g2brownianstrect}
\begin{equation}
    g_2(\tau)-1=\exp\left[-2 \left(\frac{\tau}{\tau_E}\right)^{\beta}\right] \,,
\label{eq:g2brownianstrect}
\end{equation} 
%%%%%%%%%%%%%%%%%%%%%%%%%%%%%%%%%%%%
where $\tau_E$ is the relaxation time of the coherent intermediate scattering function $f_{coh} \propto \sqrt{g_2-1}$~\cite{berne2000dynamic}. The exponent $\beta$ depends on the sample type and dynamics. For Brownian diffusion of monodisperse particles, as in Eq.~\ref{eq:g2brownian}, $\beta=1$, while for a sample exhibiting a broad distribution of relaxation times, e.g. a polydisperse suspension or a supercooled colloidal fluid, $\beta\leq 1$ (stretched exponential)~\cite{linder2002neutrons}. By contrast, compressed exponential decays, $\beta\geq 1$, are typically measured for dynamics associated to the stress relaxation of jammed materials~\cite{cipelletti2003universal}. The average relaxation time of a stretched or compressed exponential decay may be defined as 
\begin{equation}
     \tau_D = \int_{0}^{\infty}\exp\left[-2 \left(\frac{\tau}{\tau_E}\right)^{\beta}\right] \textrm{d}\tau = \frac{\tau_E}{\beta 2^{1/\beta}}\Gamma\left(\frac{1}{\beta}\right) \,,
     \label{eq:integral}
\end{equation}
with $\Gamma(x)$ the Gamma function. In particular, for the simple exponential decay case of Eq.~\ref{eq:g2brownian}, $\beta=1$ and hence $\tau_D = \tau_E/2$.

\subsection{Setup characterization}\label{sec:setupcharacterization}

As an essential step before any further measurements, we characterize our setup concerning two points that are important for the interpretation of the data presented in the following: the size of the scattering volume and the mechanical stability of the apparatus. 

Dynamic light scattering is sensitive to the relative motion of the scatterers within the scattering volume, $V_S$. In most cases, detailed knowledge of the geometry of the scattering volume is not necessary, e.g. for the Brownian particles of Sec.~\ref{sec:Brownian}, whose relative displacement is independent of $V_S$. However, there are situations where the dynamics do depend on the size and shape of $V_S$, e.g. for the beads of polymer gel of Sec.~\ref{sec:polymergel} undergoing a swelling strain $\epsilon$, where the relative motion between portions of the gel separated by a distance $L$ scales as $\epsilon L$. In our setup, the scattering volume is a thin ``cylinder'' centered around the $x$ axis. More precisely, the incoming beam cross section has a Gaussian profile, characterized by the beam waist $w_0$, the distance from the $x$ axis over which the incident intensity decays by a factor $e^{-2}$~\cite{goodman2007speckle}. In order to measure $w$ precisely, we replace the sample by a frosted glass positioned perpendicularly to the incoming beam, at $x=0$. We place a lens with focal length $f=40$ mm on the $x$ axis, right after the ground glass, and use a CMOS camera placed in the focal plane of the lens to take an image of the speckle pattern generated by the ground glass. Under these conditions, the speckle size is directly related to $w$ through the spatial auto-correlation function of the speckle pattern, $ C_I( \Delta x, \Delta y)= \langle I(x_c,y_c)I(x_c +  \Delta x ,y_c +  \Delta  y)\rangle$, with $I(x_c,y_c)$ the intensity measured at position $(x_c,y_c)$ on the detector. Indeed, for a sample illuminated by a Gaussian beam with beam waist $w_0$ and wavelength $\lambda$, the space-varying part of the auto-correlation function measured at small angles reads~\cite{goodman2007speckle}:
\begin{equation}
    C_I(\Delta x, \Delta y) \propto \exp\left[- \frac{\pi^2  w_0^2}{2 f^2 \lambda^2}(\Delta x^2+\Delta y^2)\right] \,.
\label{eq:autocorrelation}
\end{equation}
By fitting Eq.~\ref{eq:autocorrelation} to the measured $C_I(\Delta x, \Delta y)$, we find $w_0=59.5\pm0.1$ $\mu$m. 

%%%%%%%%%%%% FIG 2 %%%%%%%%%%%%%%%%%%%%%%%
\begin{figure}
\centering
\includegraphics[width=0.8\linewidth]{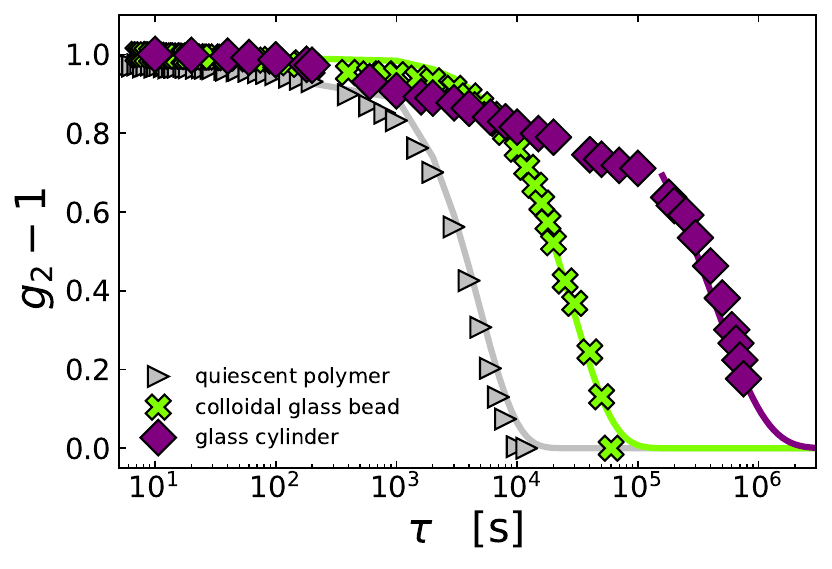}
\caption{Intensity correlation functions for a glass cylinder (purple diamonds), a colloidal glass bead (green crosses), and a quiescent bead of polymer gel  (light gray triangles). Solid lines are a fit of the slow decay to Eq.~\ref{eq:g2brownianstrect}. For the final decay of $g_2-1$, for the glass cylinder, $\tau_E=434803$ s and $\beta=1$, for the colloidal glass $\tau_E=28036$ s and $\beta=1.24$, and for the polymer gel $\tau_E=2649$ s and $\beta=1.37$.}
\label{Fig.g2stability}
\end{figure}
%%%%%%%%%%%%%%%%%%%%%%%%%%%%%%%%%%%%%%%%%

Mechanical stability is another important issue, since DLS is sensitive to displacements down to a few nanometers and the peculiar geometry of our samples makes it difficult to immobilize them as firmly as for a standard cuvette. Furthermore, samples like the gels of Sec.~\ref{sec:polymergel} may have relaxation times as long as several tens of seconds, making the requirement on mechanical stability quite stringent. We first probe the overall stability of the apparatus for a tightly fixed sample. We use a cylindrical piece of glass ($1$ cm diameter and $7$ cm height) firmly held by a clamp. The glass contains frozen impurities that scatter sufficiently for $g_2-1$ to be measurable. We plot in Fig.~\ref{Fig.g2stability} the correlation function for a ROI in the center of the cylindrical sample, measured during an experiment lasting ten days. By fitting the final relaxation of $g_2-1$ to Eq.~\ref{eq:g2brownianstrect} and using Eq.~\ref{eq:integral} we obtain a decay time $\tau_D = 217401$ s (2.3 days), which demonstrates the excellent overall mechanical stability of the apparatus.  

We then repeat stability tests for a sample configuration similar to that in experiments on spherical, water-based samples. The green crosses in Fig.~\ref{Fig.g2stability} show $g_2-1$ for a bead of colloidal glass, prepared by depositing $10\mu$L of an aqueous suspension of commercial silica nanoparticles (Ludox TM-50 from Sigma Aldrich) on a hydrophobic surface immersed in silicon oil (47 V 100, VWR chemicals). The oil has a small miscibility with water; thus, water in the suspension slowly diffuses into the oil, decreasing the drop radius from 3.94 mm to 3.31 mm. Accordingly, the particle volume fraction grows from an initial value $\varphi_i=0.31$ to $\varphi_f=0.52$, well within the glassy regime for this sample~\cite{philippe2018glass}. Since there may still be some ultra-slow relaxations in the bead of colloidal glass, the decay time obtained from a fit to the data, $\tau_D=14956$ s, provides a lower bound for the slowest dynamics that can be reliably measured on a drop sample. The compressing exponent $\beta = 1.24$ is indicative of dynamics stemming from the relaxation of internal stresses, as reported for the same system studied in the bulk~\cite{philippe2018glass} and for other jammed or glassy soft materials~\cite{duri2006length,ramos2001ultraslow,gabriel2015compressed}.

Finally, the grey triangles in Fig.~\ref{Fig.g2stability} show the $g_2-1$ function for a quiescent bead of the polymer gel investigated in Sec.~\ref{sec:polymergel}. In this case, the surrounding medium is water-saturated oil, to prevent any changes of the drop size. The measured dynamics are thus due to the spontaneous relaxation of the gel, and are expected to be slower than the driven dynamics during swelling. We find $\beta = 1.37$, larger than one, as also reported in the past for the slow dynamics of colloidal and polymer gels~\cite{cipelletti2003universal,larobina_hierarchical_2013}. The relaxation time is $\tau_D = 1460$ s, faster than that of the colloidal glass. This proves that the quiescent dynamics of the polymer gel reported here, as well as the faster dynamics during swelling discussed in Sec.~\ref{sec:polymergel}, are not affected by any artifact due to mechanical instabilities.

\section{Brownian particles in a drop} \label{sec:Brownian}
We first test our setup on a drop of a diluted suspension of Brownian particles, whose dynamics are well known and can be easily measured in a conventional DLS setup, for comparison. By running experiments that last a fraction of a second, we insure that the drop radius does not change during the measurements, regardless of the surrounding medium. Thus, the dynamics are due solely to Brownian diffusion and should be isotropic and spatially homogeneous. However, as discussed in Sec.~\ref{sec:RefractionEffects}, the relaxation time of $g_2-1$ is still expected to depend on the position of the ROI for which the intensity correlation function is calculated, due to refraction effects that lead to a spatial dependence of the scattering angle, see Fig.~\ref{fig:set-up}b. In the following, we shall demonstrate that $g_2-1$ is fully consistent with data obtained for the same sample using a conventional setup. Moreover, we will show that the space dependence of the intensity correlation function is fully accounted for by the $x$-dependence of the scattering vector $\mathbf{q}$ stemming from refraction effects.

\subsection{Materials}\label{MaterialsBrownian}

The Brownian sample is a suspension of polystyrene particles from Microparticles GmbH (diameter $2a=105$ nm), at a concentration $4\times 10^{-3}$ $\%$ w/v. The solvent is a $80/20$ w/w water/glycerol mixture, with refractive index $n=1.3585$~\cite{hoyt1934new}, viscosity  $\eta=2.3\cdot10^{-3}$ Pa s, and density $\rho = 1048~\mathrm{kg~m^{-3}}$ that nearly matches that of the particles~\cite{segur1951viscosity}. A suspension drop of volume $30$ $\mu$L (radius $R_0=1.9$ mm) is gently deposited on a hydrophobic surface (contact angle $120$ deg). To prepare the hydrophobic surface, we deposit on a microscope glass slide a 50/50 v/v water/ethanol solution of trimethoxy-(octadecyl)-silane ($20$ mg/mL). The glass slide is then heated in an oven, raising linearly the temperature $T$ from $30^{\circ}$C to $70^{\circ}$C in 1 h, followed by $12$ h at constant $T=70^{\circ}$C. To test refractive effects with opposite trends, we use either air ($n_m = 1$) or silicon oil (47 V 100 from VWR chemicals, $n_m=1.403$) as the surrounding medium. The setup has no temperature control, the room temperature is controlled and measured to be $T=22.0\pm 0.1^{\circ}$C, where the quoted uncertainty is that of the temperature probe. Sequences of speckle images lasting 60 ms are acquired using a fast camera, see Sec.~\ref{sec:optical_layout}, and processed to calculate $g_2-1$.

\subsection{Experimental results}

%%%%%%%%%%%%%%%%  FIG 3 %%%%%%%%%%%%%%%%%%%%
\begin{figure*}
\includegraphics[width=1\linewidth]{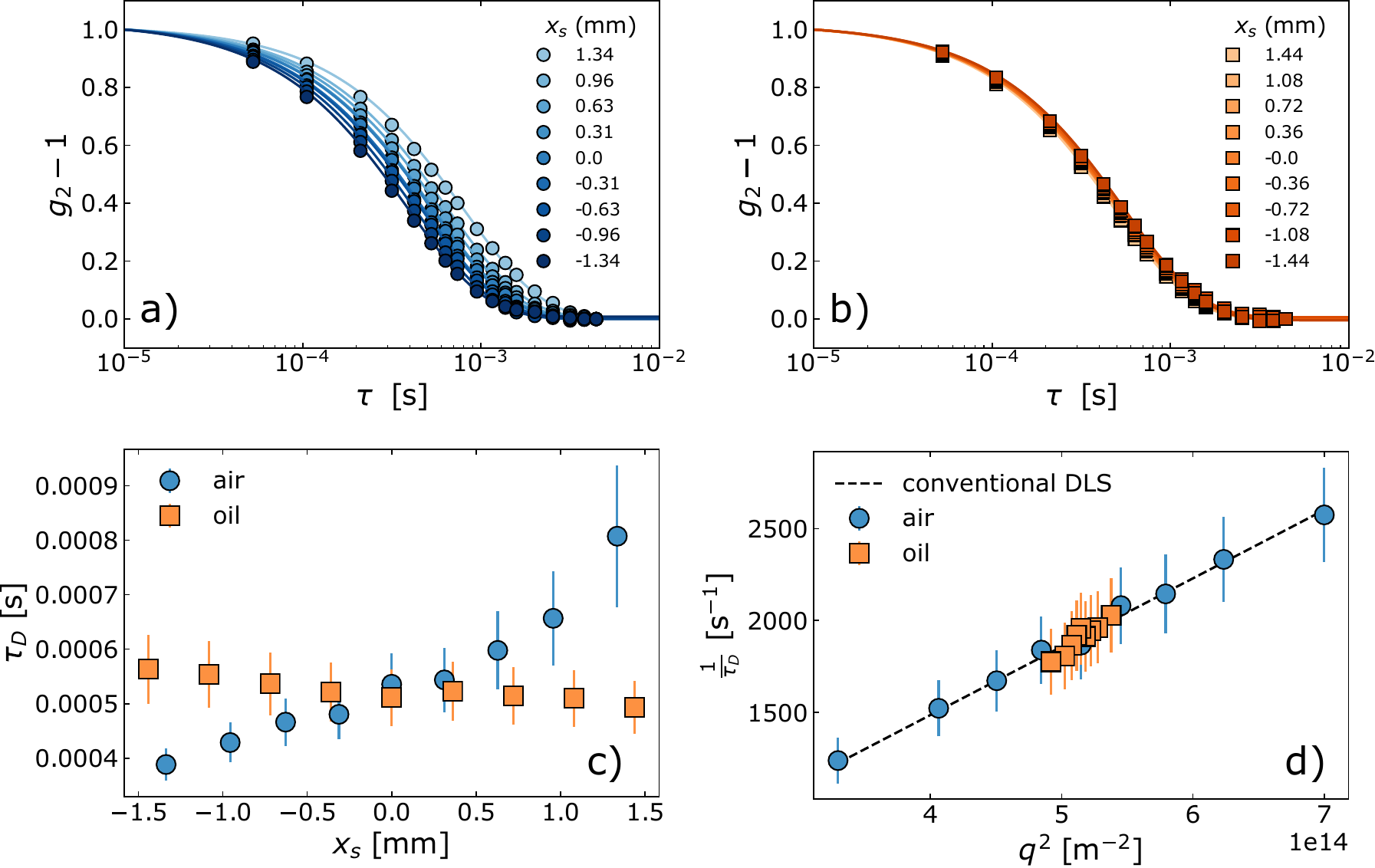}% Here is how to import EPS art
\caption{\label{fig:brownian} Intensity correlation functions for a drop of diluted suspension of Brownian particles surrounded by air (a) and oil (b), respectively. Symbols are experimental data and solid lines are fits to the data with a single exponential decay, Eq.~\ref{eq:g2brownian}. 
(c) Decay time of $g_2-1$ as a function of position inside the drop, with $x_S=0$ the center of the drop. (d) Inverse of the decay time of $g_2-1$ as a function of $q^2$. The dotted line shows $1/\tau_{D} = 2D_0q^2$, as predicted for dilute particles undergoing Brownian diffusion, with the diffusion coefficient $D_0$ independently measured by conventional dynamic light scattering at $\theta_S = 90$ deg on the same suspension, but contained in a standard DLS cuvette. In c) and d), error bars are estimated from the covariance matrix of the fit.}
\end{figure*}
%%%%%%%%%%%%%%%%%%%%%%%%%%%%%%%%%%%%%%%%%%%%

We plot in Fig.~\ref{fig:brownian}a the intensity correlation functions for a drop in air. Data are acquired for different ROIs, centered around different positions $x_S$ along a drop diameter. The correlation functions are very well fitted with a simple exponential decay as theoretically expected for monodisperse Brownian particles (Eq.~\ref{eq:g2brownian}). We find that the decay time of the correlation function, $\tau_D$, increases from $3.9$ to $8.1$ ms as $x_S$ grows. This trend is consistent with what expected from the variation of the scattering vector $\textbf{q}$ associated with the position of the ROIs for $n > n_m$. The opposite case $n < n_m$ is illustrated by the data in Fig.~\ref{fig:brownian}b, acquired for a drop immersed in oil. In this case, the relationship between $\theta_s$ and $x_S$ has an opposite trend, see  Fig.~\ref{fig:brownian}c, leading to a faster decay for smaller $x_S$. Moreover, since the refractive index mismatch between oil and water is smaller than the one between air and water, the variation of the scattering angle, and hence of $q$, is smaller for the drop in oil than for the same drop in air. This explains the smaller variation of $g_2-1$ with ROI position for the drop in oil, compare Figs.~\ref{fig:brownian}a and b. These observations are summarized in Fig.~\ref{fig:brownian}c, which shows the relaxation  time $\tau_D$ along the drop diameter for both surrounding media. To quantitatively check the relationship between position $x_S$ and $q$, we plot in Fig.~\ref{fig:brownian}d, $1/\tau_D$ as a function of $q^2$ for the data acquired with the drop in air and in oil, where $q$ is calculated from the position of each ROI as explained in Sec.~\ref{sec:RefractionEffects} and Appendix~\ref{app:qvector}. We find that the two sets of data nicely collapse on the theoretical curve $1/\tau_D = 2D_0q^2$, with $D_0$ the diffusion coefficient measured independently on the same suspension, but in the bulk, using a commercial DLS apparatus with a standard cuvette. These results confirm that the setup and data analysis presented here allow for reliably measuring the microscopic dynamics in millimeter-size drops.

\section{Swelling of a Polymer Gel}\label{sec:polymergel}

In this section we investigate a bead of a water-based polyacrylamide gel that is initially partially dried and then is swollen by immersing it in water. As for the Brownian suspension, we use sequences of speckles images to measure the microscopic dynamics. However, some caution is required when analyzing $g_2-1$ functions taken during swelling. Due to the imaging geometry, the speckle pattern associated to a given portion of the gel moves away from the center of the bead as the gel swells. This drift motion would result in a spurious fast decay of $g_2-1$. To avoid this artifact, we use the drift correction scheme of Ref.~\cite{cipelletti_simultaneous_2013a}, where the intensity correlation function is calculated in a moving reference frame that compensates for the drift. A second, more subtle effect concerns the nature of the displacements within the gel. The microscopic dynamics are due to the strain field generated during swelling, which in turn may be decomposed in a component associated to the expansion of a perfectly elastic and homogeneous spherical body, and additional displacements stemming from spatial heterogeneity in the elastic response of the gel or strain-induced rearrangements~\cite{basu2011nonaffine,aime2018microscopic}. The latter component is often the one of greater interest. Unfortunately, the geometry of our experiments does not allow one to disentangle the two components. In order to gauge the relative importance of each component, we first derive in Sec.~\ref{TheoryPolymer} a theoretical expression for the intensity correlation function during the expansion or contraction at a macroscopic strain rate $\dot{\varepsilon}$ of an ideal, perfectly elastic and homogeneous spherical body.  We then describe the preparation protocol for the gel beads in Sec.~\ref{MaterialsPolymer} and compare the experimental $g_2-1$ functions to those for the ideal case in Sec.~\ref{resultsPolymer}. Our results will be briefly discussed and compared to other studies of swelling polymer beads
in Sec.~\ref{sec:discussion_polymer}.

\subsection{Theory: intensity correlation function for the expansion or compression of a purely elastic homogeneous sphere}\label{TheoryPolymer}

%%%%%%%%%%%%%%%%  FIG 4 %%%%%%%%%%%%%%%%%%%%
\begin{figure}
\centering
\includegraphics[width=0.9\linewidth]{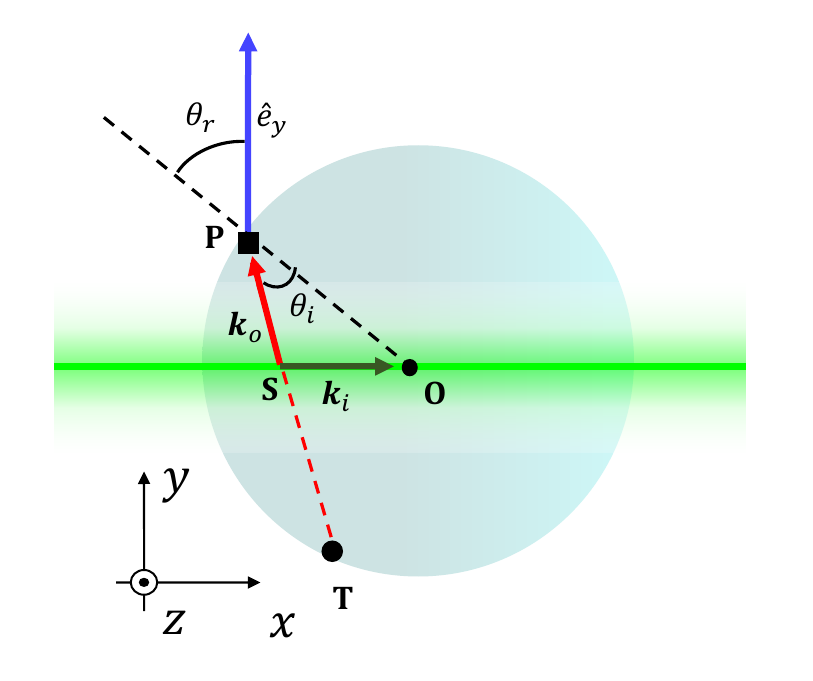}
\caption{Top view of the scattering from a water-based sample surrounded by air. The green cylinder represents the incoming laser beam, which has a Gaussian radial profile. Scattered light emerging in $P$ is issued by all points lying on the segment $PT$ (red dotted line), which is colinear with the scattered light wave vector $\textbf{k}_o$.  The blue line shows the propagation direction of the scattered light travelling outside the sphere along the $\widehat{e}_y$ direction, before being imaged on the camera sensor by the collection optics (not shown). The black dashed line shows the perpendicular to the drop surface through point $P$; $\theta_i$ and $\theta_r$ are the angles of incidence and refraction, respectively.}
\label{fig:scattering_for_theory}
\end{figure}
%%%%%%%%%%%%%%%%%%%%%%%%%%%%%%%%%%%%%%%%%%%%

We discuss here the case of a perfectly elastic and homogeneous sphere surrounded by a medium with arbitrary refractive index and undergoing expansion or contraction. We shall derive a general expression for the decay of $g_2-1$ in response to a strain increment (or decrement) $\Delta \varepsilon = \Delta R /R$, with $\Delta R$ the change of the sphere's radius. This general expression will be used in numerical calculations; a simpler analytical expression valid under some approximations will also be introduced. 

Operationally, the intensity correlation function is  calculated pixel by pixel, see Eq.~\ref{eq:cI}. Any given pixel of the CMOS detector corresponds to a small portion of the sphere surface centered around the point $P$ from which the scattered light emerges from the sphere, see Fig.~\ref{fig:scattering_for_theory}. 
The normalized intensity correlation function can be written as an integral over the contributions of all scatterers in the scattering volume $V_P$ associated to point $P$~\cite{linder2002neutrons}: 
\begin{equation}
g_2(\textbf{P},\Delta\varepsilon)-1 \propto \left|
 \int_{V_{P}} e^{-i \textbf{q}\cdot\Delta\textbf{u}(\textbf{x},\Delta\varepsilon)} I(\textbf{x},\textbf{q})\mathrm{d}^3\textbf{x}\right|^2 \,,
 \label{eq:analiticalg2_1}
\end{equation}
where $\textbf{q}$ is the scattering vector determined as shown in Sec.~\ref{sec:RefractionEffects}, $I(\textbf{x},\textbf{q})$ is the scattered intensity originating from a point at position $\mathbf{x} = x\hat{e}_x + y\hat{e}_y + z\hat{e}_z$ and $\Delta\textbf{u}(\textbf{x},\Delta\varepsilon)$ is the displacement field generated by the strain increment $\Delta\varepsilon$. Note that in writing Eq.~\ref{eq:analiticalg2_1} we have dropped inessential prefactors, keeping in mind that the final expression for the correlation function has to be normalized such that $g_2-1 \rightarrow 1$ for $\Delta \varepsilon \rightarrow 0$.

In our setup, the pixel size corresponds roughly to the speckle size, and both are $<<R$. Given that the intensity value does not change significantly within a speckle~\cite{goodman2007speckle}, the integral in Eq.~\ref{eq:analiticalg2_1} may be simplified by recognising that $I$ and $\Delta \mathbf{u}$ depend essentially only on position along the segment $PT$. Thus, the 3D integration over $V_P$ may by replaced by a one-dimensional integration over the scalar $l\in [0,1]$ that parametrizes the position along $PT$:
\begin{equation}
    \textbf{x}(l)=l(\textbf{OP}-\textbf{OT}) +\textbf{OT} \,,
    \label{eq:pointinthesphere}
\end{equation}
with $\textbf{x}(0)=\textbf{OT}$ and $\textbf{x}(1)=\textbf{OP}$ (see Appendix~\ref{app:analyticalg2detal} for expressions of the coordinates of $P$ and $T$). Furthermore, we note that the scattered intensity $I(\mathbf{x},\mathbf{q})$ is proportional to $I_0(\mathbf{x})$, the $\mathbf{x}$-dependent intensity of the incident beam. The proportionality factor accounts for the $q$ dependence of the scattered light; in this context, this proportionality factor is unimportant, since it may be included in the normalization of $g_2-1$. Given that the incident beam has a Gaussian profile and dropping again inessential prefactors, one has
\begin{equation}
    I_0[\textbf{x}(l)] \propto \exp\left\{-\frac{ 2 \left[y(l)^2+z(l)^2\right]}{w_0^2}\right\}\,.
    \label{eq:intesity0}
\end{equation}
%NOTE: I've dropped \left[\frac{w_0}{w(x)}\right]^2
In writing Eq.~\ref{eq:intesity0}, we have used the fact that the incoming beam is mildly focused, such that the Rayleigh range is comparable to the drop diameter and the cross section $w$ of the beam is nearly constant and equal to the beam waist $w_0$. With these assumptions, Eq.~\ref{eq:analiticalg2_1} simplifies to
\begin{multline}
g_2(\textbf{P},\Delta\varepsilon)-1 \propto \\
\left|
 \int_{0}^1 e^{-i \textbf{q}\cdot\Delta\textbf{u}[\textbf{x}(l),\Delta\varepsilon]} \exp\left\{-\frac{ 2 \left[y(l)^2+z(l)^2\right]}{w_0^2}\right\}\mathrm{d}l\right|^2 \,.
 \label{eq:analiticalg2_2}
\end{multline}

The displacement field for the compression or expansion of a homogeneous, purely elastic sphere reads~\cite{bower2009applied}
\begin{equation}
    \begin{cases}\Delta\textbf{u}(\mathbf{x},\Delta\varepsilon) = \left[ C_1(\Delta\varepsilon) r+ \frac{C_2(\Delta\varepsilon)}{r}\right] \hat{r}  \,, r \ge r_c\\
    \Delta\textbf{u}(\mathbf{x},\Delta\varepsilon) = 0 \,, r < r_c
    \end{cases}
    \label{eq:displacementfield}
\end{equation}
with $r = |\mathbf{x}|$, $\hat{r} = \mathbf{x}/r$, $r_c$ the radius of an incompressible core, and $C_1$, $C_2$ two constants such that
\begin{equation}
\label{eq:disp_C}
    \begin{cases}
      C_1 = -\frac{ C_2}{r_c^{3}}\\
      C_2 = R_i \Delta\varepsilon \left( -\frac{R_i}{r_c^{3}} +  \frac{1}{R_i^{2}}\right)
    \end{cases}\
\end{equation}
where $R_i$ is the initial radius of the sphere. Mathematically, $r_c$ is introduced to avoid the divergence of $\Delta \mathbf{u}$ at the sphere center; physically, is accounts for the fact that the continuum medium description of Eq.~\ref{eq:displacementfield} must break down at the structural length scale of the gel. In practice, we expect $r_c$ to be of the order of the monomer size or the mesh size at most; we checked that varying $r_c$ in the range $10^{-9}$---$10^{-5}$ m, $\Delta \mathbf{u}$ and hence $g_2-1$ do not vary appreciably. 

Equation~\ref{eq:analiticalg2_2}, together with the expressions for the displacement field, Eqs.~\ref{eq:displacementfield}-\ref{eq:disp_C}, may be solved numerically to obtain the decay of $g_2-1$ as a function of the strain increment $\Delta \varepsilon$ or, assuming a constant expansion or contraction rate $\dot{\varepsilon}$, for increasing time lags $\tau = \Delta\varepsilon / \dot{\varepsilon}$. Note that $g_2-1$ is insensitive to the sign of $\Delta \varepsilon$, i.e. the same correlation function is recovered for compression or expansion, provided that $|\Delta \varepsilon|$ is the same.

Equation~\ref{eq:analiticalg2_2} holds quite generally and may be used to compute $g_2-1$ for various kinds of dynamics, using the suitable expression for $\Delta \mathbf{u}$. For the case of expansion or contraction discussed here, we show in Appendix~\ref{app:analyticalg2detal} that an approximated analytical solution can be found for Eq.~\ref{eq:analiticalg2_2}, provided that one uses a simplified expression for the displacement field, dropping the $1/r$ term in Eq.~\ref{eq:displacementfield} and assuming 
\begin{equation}
    \Delta\textbf{u}(\textbf{x},\Delta\varepsilon) = \textbf{x}\Delta\varepsilon \,.
    \label{eq:simplifyeddisplacementfield}
\end{equation}
Under this hypothesis, we show in Appendix~\ref{app:analyticalg2detal} that the correlation function decays as a Gaussian function:
\begin{equation}
    g_2(\tau)-1=\exp\left[-2 \left(\frac{\tau}{\tau_E}\right)^2\right] \,,
\label{eq:g2analytical_simplified}
\end{equation} 
where the relaxation time $\tau_E$ depends on the position of point $P$, the strain rate $\dot{\varepsilon}$, the scattering vector $\mathbf{q}$ and the beam waist $w_0$, see Eq.~\ref{eq:tauE}.

%%%%%%%%%%%%%%%%  FIG 5 %%%%%%%%%%%%%%%%%%%%
\begin{figure}
\centering
\includegraphics[width=0.8\linewidth]{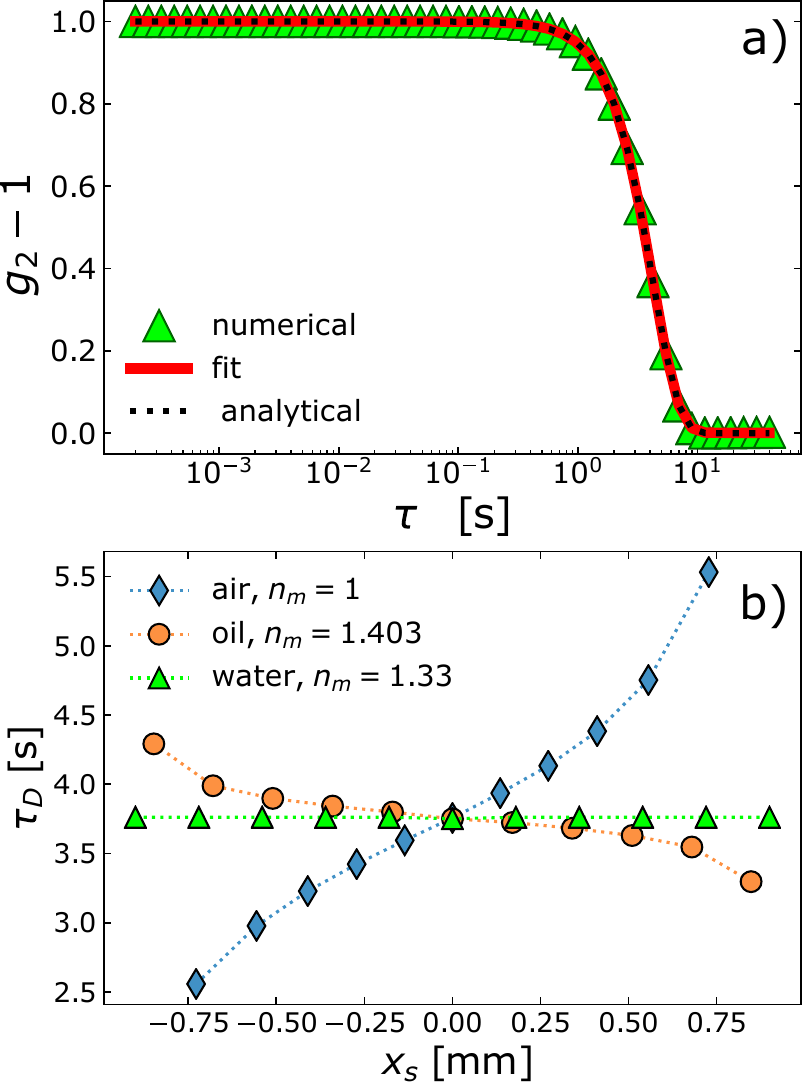}
\caption{a) Symbols: intensity correlation function calculated numerically for a water-based spherical bead of gel in water undergoing contraction or expansion at a fixed strain rate, see text for details. The red solid line is a fit to the data using Eq.~\ref{eq:g2brownianstrect}, yielding $\tau_E = 6.15$ s and $\beta=2$. The black dotted line is $g_2-1$ obtained via the analytical approximation of Eq.~\ref{eq:g2analytical_simplified}. b) Decay time $\tau_D$ as a function of position obtained from the numerical $g_2-1$, for different surrounding media, as indicated by the lablel.}
\label{fig:g2timesim}
\end{figure}
%%%%%%%%%%%%%%%%%%%%%%%%%%%%%%%%%%%%%%%%%%%%

The triangles in Fig.~\ref{fig:g2timesim}a show an example of the intensity correlation function $g_2(\textbf{P},\tau)-1$ obtained numerically using the laser beam parameters of our setup, for a point $\textbf{P}=(0.001,0.9999,0.001)~\mathrm{mm}$ on a sphere with $R=1$ mm, refractive index $n=1.33$, and surrounded by water, $n_m=1.33$. The strain increment varies in the range $10^{-4} \le \Delta \varepsilon \le 6.3 \times 10^{-3}$ and $g_2-1$ is plotted as a function of time delay $\tau$, assuming expansion or contraction at a constant rate $\dot{\varepsilon}=5\times 10^{-4}~\mathrm{s}^{-1}$. The continuous line shows a fit to the the numerical data with Eq.~\ref{eq:g2brownianstrect}, yielding $\tau_E = 6.15$ s and $\beta=2$. The dotted line shows the approximated analytical solution, Eq.~\ref{eq:g2analytical_simplified}, where the relaxation time $\tau_E$ has been calculated based on the same parameters as for the numerical calculations, using Eq.~\ref{eq:tauE}. We find that the approximated analytical form is indistinguishable from both the numerical calculation and its fit, indicating that for the typical experimental parameters of our measurements the simplified expression of the displacement field, Eq.~\ref{eq:simplifyeddisplacementfield}, is in excellent agreement with the full expression.

Finally, to demonstrate the impact on $g_2-1$ of the refractive index mismatch between the gel bead and the surrounding medium, we show in Fig.~\ref{fig:g2timesim}b the decay time $\tau_D$ obtained from the numerical $g_2-1$ as a function of the position in the bead, for different surrounding media: air, water and oil. In the center of the bead ($x_S=0$), we obtain the same decay time independently of $n_m$, as expected since in this position the scattering angle, and hence $q$, does not depend on $n_m$, see Fig.~\ref{fig:set-up}b. For $x_S \ne 0$, $\tau_D$ depends in general on the refractive index mismatch. For $n_m = n=1.33$, which is the case of hydrogel beads immersed in water (green triangles in Fig.~\ref{fig:g2timesim}b), $\tau_D$ is independent of position. By contrast, for a sample surrounded by air, $n_m=1$, the decay time varies by more than a factor of two along the bead diameter, reflecting the change in $q$. For a sphere surrounded by oil, $n_m=1.403$, a reverse trend of $\tau_D(x_S)$ is seen (compare the blue diamonds and the orange circles), since $n_m-n$ has an opposite sign compared to the case of a bead in air. Note also that the variation of $\tau_D$ is smaller, since the refractive index mismatch is lesser.

\subsection{Experiments: Materials}\label{MaterialsPolymer}

Polyacrylamide gels are prepared by copolymerization of acrylamide (46.03 g/L) as a monomer and methylenebisacrylamide (0.373 g/L) as a co-monomer in the presence of tetramethylenediamine (0.6 g/L) and sodium persulfate (0.93 g/L) as initiators in water. All chemicals are quickly mixed together and a 30 $\mu$L drop of the solution is injected in an Petri dish, filled with silicon oil (47V 100 from VWR chemicals), where polymerization occurs. The oil has been previously saturated with nitrogen, since polymerization is inhibited by oxygen. The oil mass density is $97~\mathrm{kg}~\mathrm{m}^{-3}$ at 25 °C, nearly equal to that of water, such that gravity effects are negligible and the drop has a spherical shape, due to surface tension. Changing the relative amounts of the monomer and co-monomer allows one to tune the gel mechanical properties; the beads used here have an elastic modulus $G_0=700$ Pa ~\cite{arora2018impact}. The mesh size of the gel, defined as the average distance between two crosslinks and estimated from mechanical measurements, is $\zeta\approx(k_B T/G_0)^{\frac{1}{3}}\approx 20$ nm. 

The gel bead is first shrunk by letting it dry in air for about 1 hour. During this phase, the bead rests on a super-hydrophobic substrate prepared as described in Sec.~\ref{MaterialsBrownian} and placed in the sample chamber left open, resulting in a $\sim 14\%$ decrease of the bead radius. The sample chamber, an optical glass box of size $5 \times 5 \times 5~\mathrm{cm}^3$, is then filled with water, initiating swelling and defining the experiment time $t=0$. All measurements are performed in a room where  $T = 22.0 \pm 0.1^{\circ}$C. In order to keep illuminating the bead in its equatorial plane all along the swelling process, the bead is  translated upwards using a micrometric vertical translation stage in between the acquisition of series of speckle images.

\subsection{Experimental results}\label{resultsPolymer}

\begin{figure}
\centering
\includegraphics[width=1\linewidth]{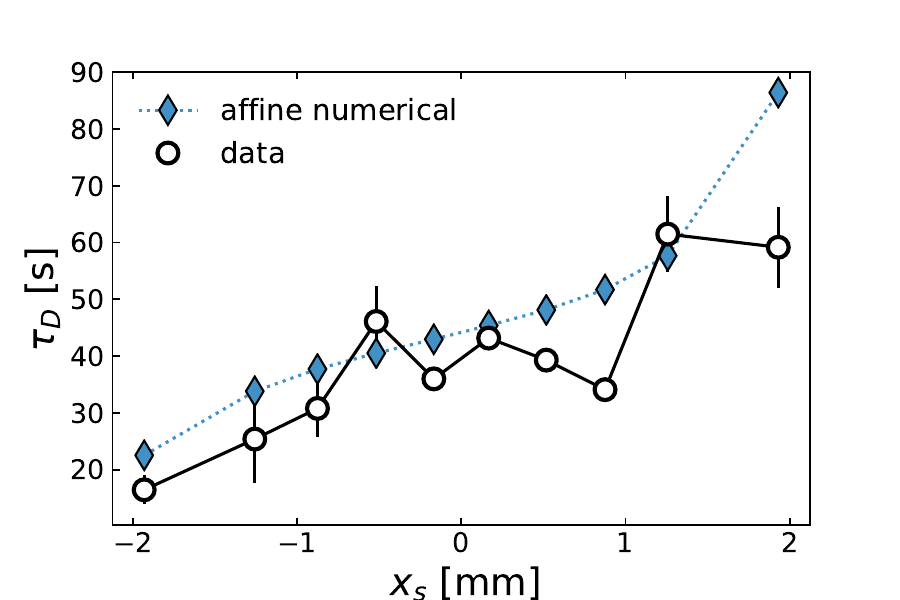}
\caption{Decay time $\tau_D$ as a function of position for a bead of polymer gel drying at a rate $\dot{\epsilon}=4.3\times 10^{-5}$ s$^{-1}$. Empty circles are experimental data, blue diamonds are numerical results obtained using Eqs.~\ref{eq:analiticalg2_2}-\ref{eq:disp_C} for a homogeneous, purely elastic sphere undergoing compression at the same rate as in the experiments.}
\label{fig:figdrying}
\end{figure}

We show in Fig.~\ref{fig:figdrying} the microscopic relaxation time measured during the last 14 minutes of the drying phase that precedes the swelling experiment. Remarkably, we find that $\tau_D$ is in very good agreement with both the spatial trend and the magnitude predicted for the microscopic relaxation time for a purely elastic sphere undergoing isotropic contraction. In the following, for the sake of simplicity, we shall refer to the radial displacement field associated with the contraction or expansion of such a homogeneous, purely elastic sphere as to the ``affine dynamics'', and shall term ``non-affine dynamics'' any microscopic displacements that may add up to the affine component. Note that, although strictly speaking the strain field for an expanding or contracting sphere is not purely affine due to the $1/r$ term in the r.h.s. of Eq.~\ref{eq:displacementfield}, in practice in our experiments this term is negligible, justifying this terminology.  The data in Fig.~\ref{fig:figdrying} thus demonstrate that the microscopic dynamics during the late stages of drying are essentially purely affine, with a local strain rate that is identical to the macroscopic one, for all probed $x_S$. This suggests that any radial gradients in solvent content that may have developed in the initial stages of drying have relaxed by the time the swelling experiment starts.   

%%%%%%%%%%%%%%%%  FIG 6 %%%%%%%%%%%%%%%%%%%%
%%%%%%%%%%%%%%%%  END FIG 6 %%%%%%%%%%%%%%%%%%%%

%Figure~\ref{fig:imaging_polymer} shows representative images illustrating the gel swelling at the macroscopic level.  %The microscopic dynamics are measured throughout the swelling using the photon correlation imaging setup described in Sec.~\ref{Materialsandmethods}. 

\begin{figure*}
\includegraphics[width=1\linewidth]{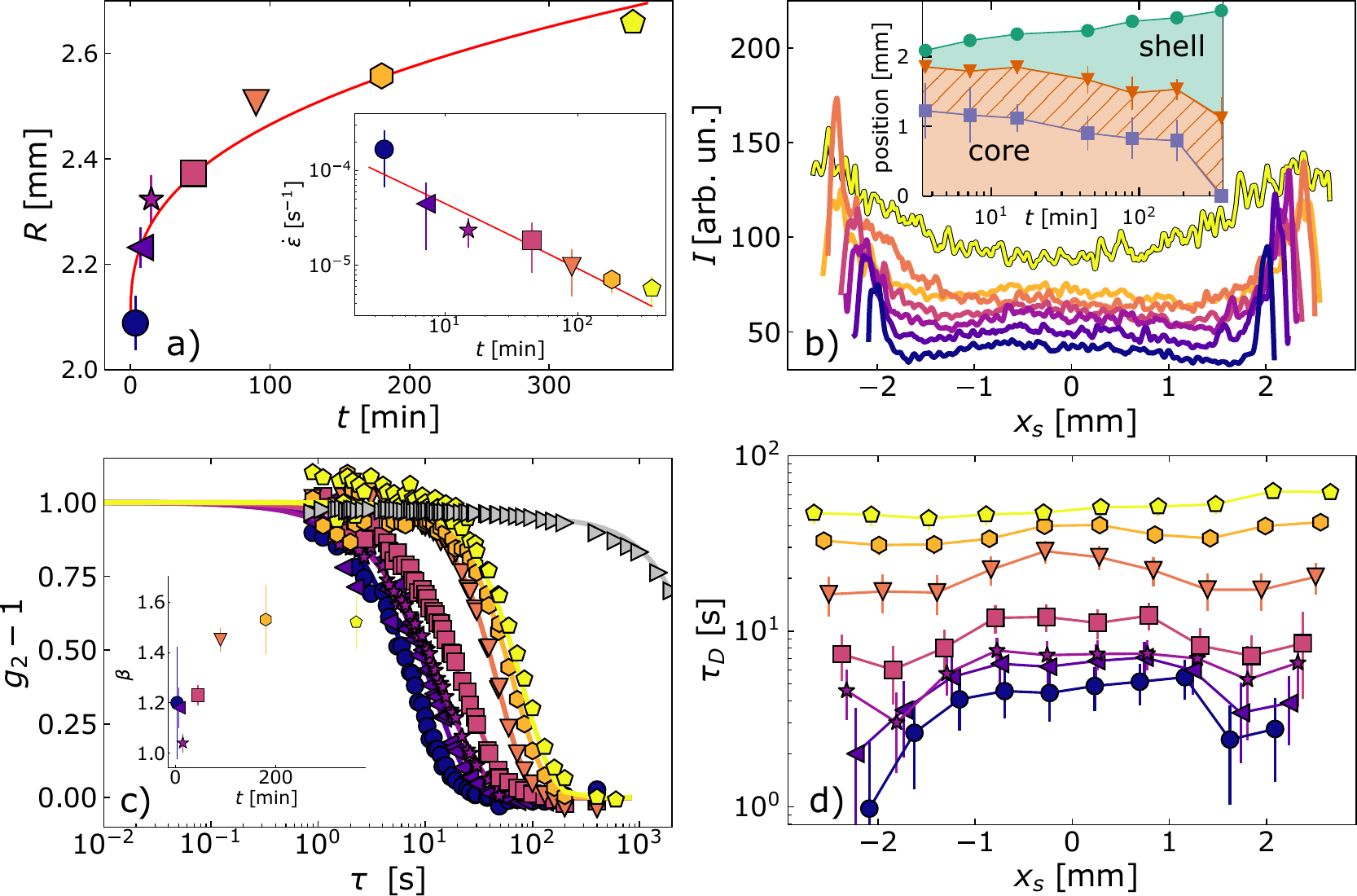}% Here is how to import EPS art
\caption{\label{fig:micro_swelling}(a) Radius of a swelling polyacrylamide gel bead as function of time. The line is a power law fit to the increment of $R$, see the text for details. Inset: strain rate as a function of time, measured directly from the speckle images (symbols) or by differentiating the power law fit of the main panel (line). Error bars represent the uncertainty of the gel boundary as measured from the speckle images. (b) Scattered intensity as a function of position for various $t$, indicative of the formation of a core-shell structure upon swelling. For clarity, curves have been offset vertically in increments of 5 gray units. From bottom to top, $t$ = 3.6, 7.5, 15, 45, 90, 180, and 360 min, same colors as in a). The inset shows the time evolution of the bead radius (circles) and of the core-shell interface, as inferred from the $I(x_S)$ data shown in the main graph (triangles), or from the relaxation time of panel d) (squares), see text for details. (c) Intensity correlation functions measured during the swelling of the bead, for a ROI located at the center of the bead, $x_S = 0$, same $t$ values and color code as in b). The solid lines are fits with Eq.~\ref{eq:g2brownianstrect}, yielding $\tau_D=$ 4.44, 6.52, 7.25, 12.04, 27.94, 39.86, 47.34 s and $\beta=$ 1.20, 1.18, 1.04, 1.23, 1.45, 1.53, 1.52 for increasing $t$. The grey right triangles show $g_2-1$ for a gel where swelling is prevented, see Sec.~\ref{sec:setupcharacterization} (same data as in Fig.~\ref{Fig.g2stability}). Inset: time dependence of the stretching exponent $\beta$. (d) Decay time $\tau_D$ as a function of position $x_S$ along the bead diameter, same $t$ values, symbols and colors as in a), b) and c). }
\end{figure*}

Figure~\ref{fig:micro_swelling} describes the gel behavior during the swelling phase. The time evolution of the radius of the gel bead as obtained from the speckle images is shown in Fig.~\ref{fig:micro_swelling}a. At $t=0$ the bead radius is $R_{0}=2.08$ mm; as the gel absorbs water and swells, $R$ grows reaching $R_{\rm{f}}=2.63$ mm at the end of the experiment, corresponding to a maximum engineering strain $\varepsilon_{\rm{sw}}=(R_{\rm{f}}-R_{0})/R_{0}=28\%$. During swelling, the radius growth can be very well fitted by $R(t) = R_0 + C t^{\gamma}$, with $R_0$ fixed to $2.08~\rm{mm}$, and fitting parameters $C = 0.076 \pm 0.022~\rm{mm}~\rm{min}^{-\gamma}$ and $\gamma = 0.35 \pm 0.06$ fitting parameters. The macroscopic swelling rate $\dot \varepsilon = R(t)^{-1}\mathrm{d}R/\mathrm{d}t$ obtained from the fit to $R(t)$ is shown as a red line in the inset of Fig.~\ref{fig:micro_swelling}a: it is essentially indistinguishable from a power law decay $\dot \varepsilon \propto t^{-0.69}$ and is in very good agreement with the swelling ratio obtained at selected $t$ by measuring the increment of $R$ over a time interval $\Delta t$ directly from the speckle images, using $\Delta t \approx 3.5~\mathrm{min}$ for the first two points and $\Delta t \approx 14~\mathrm{min}$ for $t\ge 15~\mathrm{min}$.

Figure~\ref{fig:micro_swelling}b shows the spatial dependence of the scattered intensity $I$, for various $t$. Data are obtained by averaging $I(x,z)$ along the $z$ (vertical) direction and over a sliding window of three pixels in the $x$ direction. We find that $I$ is not uniform, revealing time-dependent spatial heterogeneity in the gel structure and density. For a polymer gel, the scattered intensity at a given $q$ depends on material parameters such as the monomer size and refractive index, as well as the monomers concentration and their spatial structure. This results in a complex behavior, such that in general $I(q)$ is not simply proportional to monomer concentration and may even decrease with increasing concentration. This is indeed the case for our gel: by calculating the total scattered intensity, $I_{tot}(t) = \int I(x_S,t) \mathrm{d}x_S$, we find that $I$ increases as the gel swells, i.e. as the average monomer concentration decreases (see Fig.~\ref{fig:Igel}a in Appendix~\ref{app:Igel}). The high-intensity regions seen at the bead edges in Fig.\ref{fig:micro_swelling}b thus correspond to a shell of swollen gel that initially forms at the gel-solvent interface and then progressively propagates inwardly. We determine the position of the shell-core interface by analyzing the spatial dependence of $I$ (see Appendix~\ref{app:Igel} and Fig.~\ref{fig:Igel}b for details) and plot in the inset of Fig.~\ref{fig:micro_swelling}b its time dependence, together with that of the bead radius. We find that the shell-core structure tends to fade as the swollen region invades the whole bead.

%%%%%%%%%%%%%%%%  FIG 7 %%%%%%%%%%%%%%%%%%%% ~\ref{fig:micro_swelling}
%%%%%%%%%%%%%%%% END FIG 7 %%%%%%%%%%%%%%%%%%%%%%%%

To gain insight on the swelling phenomenon at the microscopic level, we inspect correlation functions measured at various $t$ and positions in the gel. Since the hydrogel is immersed in water, the solvent refractive index is the same of that of the surrounding medium, $n=n_m$. Thus, there are no refractive effects: regardless of $x_S$, the dynamics are probed at the same scattering vector $\textbf{q}$, corresponding to a typical length scale $1/q\approx 45$ nm, comparable to a few mesh sizes. 

In Fig.~\ref{fig:micro_swelling}b we plot the intensity correlation functions for a region in the center of the bead, $x_S=0$, and various $t$. During swelling, the correlations functions decay much faster than for a quiescent gel, compare the colored symbols to the grey right triangles. This indicates that, due to the interconnected structure of the gel, stresses due to swelling generate strain fields and microscopic rearrangements throughout the whole network, even at the beginning of the experiment, where swelling occurs only in a thin shell at the gel periphery, as shown in Fig.~\ref{fig:micro_swelling}b. Over the duration of experiment ($6$ hours), the dynamics slow down by more than one order of magnitude, mirroring the decrease of the macroscopic expansion rate $\dot{\varepsilon}$ shown in the inset of Fig.~\ref{fig:micro_swelling}a. We quantify the microscopic dynamics by fitting $g_2 -1$ to a stretched exponential function, Eq.~\ref{eq:g2brownianstrect}, expressing the microscopic decay time as $\tau_D$, calculated using Eq.~\ref{eq:integral}. We find that the decay time $\tau_D$ increases with time by an order of magnitude, ranging from $4.44$ s at the beginning of the swelling to $47.34$ s at $t=360$ min. Concomitantly, the stretching exponent evolves from $\beta \approx1$ in the early stages of swelling to $\beta \approx 1.5$ for $t \ge 90$ min (inset of Fig.~\ref{fig:micro_swelling}b). 

We find that the microscopic dynamics have a non-trivial spatio-temporal dependence. Figure~\ref{fig:micro_swelling}c displays the characteristic relaxation time $\tau_D$ at different times during the swelling process, as a function of position $x_S$ along the bead diameter. The slowing down of the microscopic dynamics as the gel swells, observed in the center of the bead, is seen at all $x_S$. However, at the beginning of the swelling process the microscopic dynamics clearly depend on position: 
$\tau_D$ is roughly two times smaller near the bead edges as compared to in the center, consistent with the notion of a shell that swells much faster than the bead core. This space dependence becomes weaker with time, and eventually vanishes: nearly uniform microscopic dynamics are recovered $180$ min from the beginning of the experiment. Remarkably, we find that there is a difference in the shell-core structure as inferred from static and dynamic measurements, Figs~\ref{fig:micro_swelling}b and c,d, respectively. To quantify this difference, we apply a procedure similar to that used for the $I(x_S)$ data to the space-dependent relaxation time of Fig.~\ref{fig:micro_swelling}d, determining the position of the core-shell interface as inferred from measurements of the dynamics. The results are shown in the inset of Fig.~\ref{fig:micro_swelling}b (square symbols). The core inferred from the dynamics is consistently smaller than that obtained from the static $I(x_S)$, and vanishes at $t=360$ min, when no core-shell structure can be distinguished in the spatial dependence of the dynamics.

%%%%%%%%%%%%%%%%  FIG 8 %%%%%%%%%%%%%%%%%%%% ~\ref{fig:non_affine}
\begin{figure}
\includegraphics[width=1\linewidth]{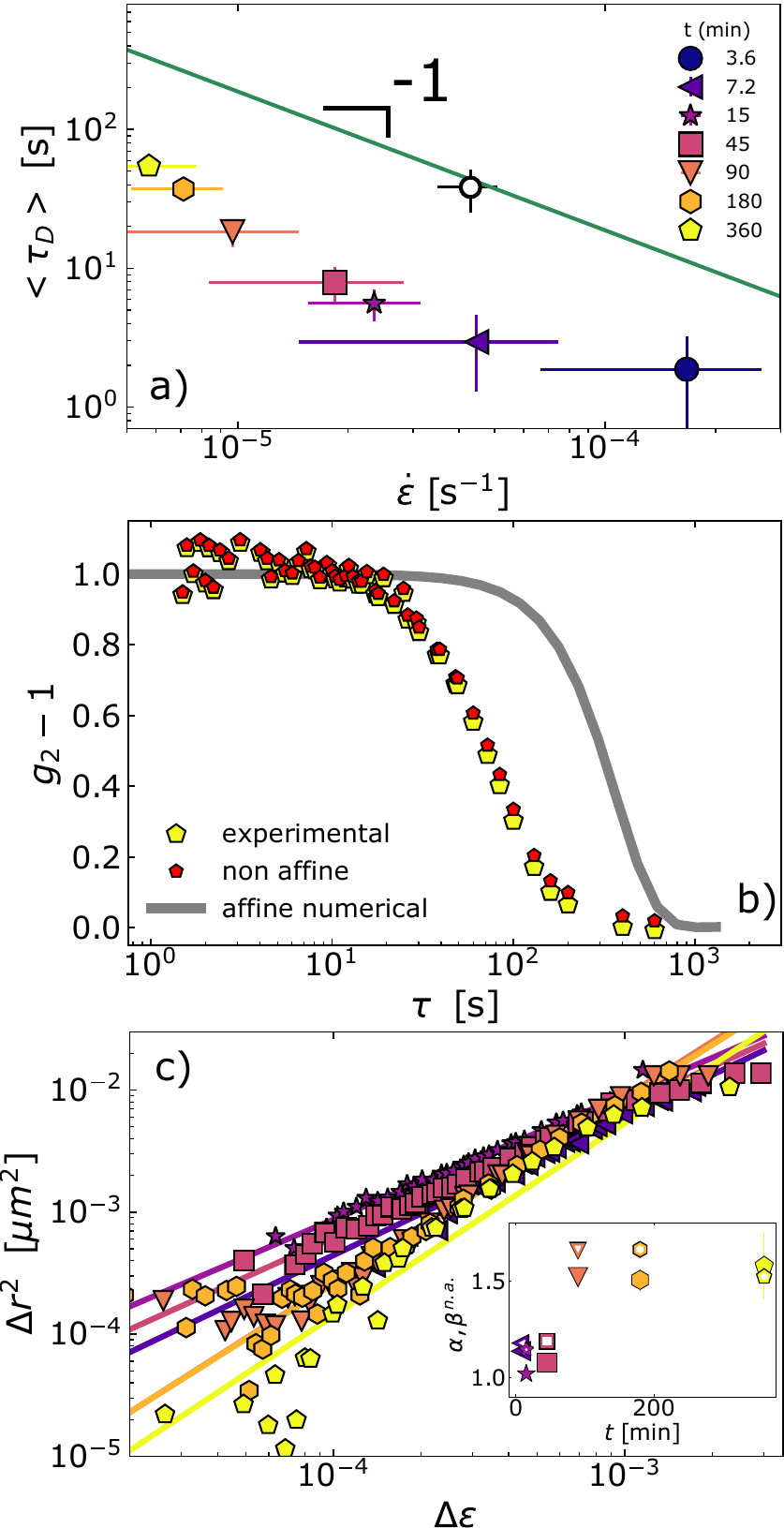}
\caption{\label{fig:non_affine} (a) Decay time $\tau_D$ averaged over position $x_S$, as a function of strain rate $\dot{\varepsilon}$. Error bars on $<\tau_D>$ are the standard deviation of the data shown, for each $t$, in Fig.~\ref{fig:micro_swelling}d. The green line is the decay time expected for the expansion of a purely elastic, homogeneous sphere, as obtained by fitting correlation functions calculated numerically by inserting the experimental $\dot \varepsilon$ in Eq.~\ref{eq:analiticalg2_2}. The empty circle is $<\tau_D>$ for the shrinking gel measured during drying, obtained form the data of Fig.~\ref{fig:figdrying}. (b) Intensity correlation functions for a polymer gels swelling at rate $\dot{\varepsilon}=5.73\times 10^{-6}~\mathrm{s}^{-1}$. Gray solid line: theoretical intensity correlation function for purely affine dynamics; large yellow symbols: experimental $g_2-1$ for $x_S=0$;  small red symbols: corresponding intensity correlation function for the non-affine displacements only, calculated using Eq.~\ref{eq:affine_and_non_affine}. (c) Non-affine mean square displacement $\Delta r^2$ for $x_S=0$, obtained via Eq.~\ref{eq:nonaffineg2} as a function of the strain increment $\Delta\varepsilon$ during swelling. The solid lines are power law fits, $\Delta r^2 \propto \Delta\varepsilon^{\alpha}$, the exponent $\alpha$ is shown in the inset (solid symbols), together with $\beta^{n.a.}$ (open symbols), the stretching exponent obtained by fitting $g_2^{n.a.}-1$ with a stretched exponential. In both the main plot and the inset the symbols are as in panel a).}
\end{figure}
%%%%%%%%%%%%%%%%%%%%%%%%%%%%%%%%%%%%%%%%%%%%

As discussed at the beginning of Sec.~\ref{sec:polymergel}, our setup does not allow one to separately measure the dynamics along the radial direction and perpendicularly to it, which makes it difficult to directly assess the relative importance of affine and non-affine dynamics. The data shown in Figs.~\ref{fig:micro_swelling}c,d, however, already point to deviations from affinity: indeed, we recall that for affine dynamics $\beta=2$ as opposed to $\beta \lesssim 1.5$ as measured here, and that for $n = n_m$ $\tau_D$ should be independent of position (see the triangles in Fig.~\ref{fig:g2timesim}b), which is not the case in our experiments, at least in the initial stages of swelling, see Fig.~\ref{fig:micro_swelling}d. To further investigate deviations from affinity, we compare $\tau_D$ as obtained in the swelling experiment to the theoretical relaxation time for purely affine dynamics, which we calculate using Eqs.~\ref{eq:analiticalg2_2}-\ref{eq:disp_C} with $\Delta \varepsilon = \tau \dot \varepsilon $, where $\dot \varepsilon$ is the experimental macroscopic strain rate and $\tau$ the experimental time delay of $g_2-1$. Figure~\ref{fig:non_affine}a compares the theoretical relaxation time of affine dynamics (line) to the experimental relaxation time $<\tau_D>$, obtained by averaging over space $\tau_D$. The experimental dynamics are up to a factor of ten faster than the theoretical prediction. Furthermore, for affine motion $\tau_D$ is inversely proportional to $\dot{\varepsilon}$, see the line in Fig.~\ref{fig:non_affine}a and Eq.~\ref{eq:tauE}, while the experimental $<\tau_D>$ clearly deviates from such a $\dot{\varepsilon}^{-1}$ dependence.

These discrepancies indicate that additional non-affine relaxations occur during the swelling process. In order to quantify non-affine dynamics, we assume that the affine and non-affine contributions to $g_2-1$ are uncorrelated. Accordingly, the intensity correlation function may be factorized:
\begin{equation}
    g_2(\Delta\varepsilon,\textbf{q}) -1 = \left [ g_2^{aff.}(\Delta\varepsilon,\textbf{q}) -1 \right ] \left [ g_2^{n.a.}(\Delta\varepsilon,\textbf{q}) -1\right ] \,,
\label{eq:affine_and_non_affine}
\end{equation}
where the superscripts $aff.$ and $n.a.$ indicate affine and non-affine contributions, respectively. By dividing the experimental $g_2-1$ by the affine correlation function $g_2^{aff.}(\Delta\varepsilon,\textbf{q})-1$ obtained numerically using the macroscopic strain rate $\dot \varepsilon$, we obtain the non-affine correlation function. Figure~\ref{fig:non_affine}b shows an example for $\dot \varepsilon = 5.74\times 10^{-6}~\mathrm{s}^{-1}$, the strain rate for which the experimental and theoretical decay times are closest and thus the experimental $g_2-1$ closest to the affine contribution. Even in this case, the experimental correlation function is in fact very close to its non-affine contribution, reflecting the fact that non-affine dynamics always dominate the relaxation of $g_2-1$, as already inferred from the comparison of the relaxation times shown in Fig.~\ref{fig:non_affine}a. Note that this is in sharp contrast to the behavior during the drying stage (open circle in Fig.~\ref{fig:non_affine}a).

A thorough interpretation of the non-affine dynamics would require collecting data at different $q$ vectors, which would allow one, e.g., to unambiguously distinguish between diffusive and ballistic dynamics. Unfortunately, this is not possible with our setup, due to the limitations imposed by the spherical shape of the gel beads. To make progress, we assume that, for any given $\Delta \varepsilon$ (or, equivalently, any $\tau$), the probability distribution of the non-affine displacements is Gaussian. This hypothesis is motivated by previous work on polyacrylamide gels seeded with micron-sized tracer particles and submitted to a shear strain~\cite{basu2011nonaffine}. The particle displacements were found to have both an affine and a non-affine component, the latter being distributed according to a Gaussian function. Under this assumption, the intensity correlation function depends on the (non-affine) particle mean square displacement $\Delta r^2 (\Delta \varepsilon)$ as
\begin{equation}
    g_2^{n.a.}(\Delta\varepsilon,\textbf{q}) -1 = \exp\left[-2 (\Delta r^2q^2)/3\right] \,,
    \label{eq:nonaffineg2}
\end{equation}
a result easily obtained from standard dynamic light scattering theory~\cite{berne2000dynamic,puseyDLSchapter} by replacing the usual time delay $\tau$ by $\Delta \varepsilon = \tau \dot\varepsilon$.

We show in the double-logarithmic plot of Fig.~\ref{fig:non_affine}c the non-affine mean square displacement $\Delta r^2(\Delta \varepsilon)$  for $x_S=0$, obtained from Eq.~\ref{eq:nonaffineg2} using $g_2^{n.a.}-1$ calculated according to Eq.~\ref{eq:affine_and_non_affine} (data for $t=3.6$ min are omitted because they are too noisy). The lines are power law fits to the data, yielding an exponent $\alpha$, whose time dependence is shown by the solid symbols in the inset. In the initial stages of swelling, $\alpha \approx 1$, indicative of diffusive dynamics, consistent with the shape of $g_2^{n.a.}-1$, for which a fit with Eq.~\ref{eq:g2brownianstrect} yields an exponent $\beta^{n.a.}$ close to $1$, as expected for a diffusive process~\cite{berne2000dynamic,puseyDLSchapter} (see the open symbols in the inset of Fig.~\ref{fig:non_affine}c).
%~\footnote{The values of $\beta$ shown in Fig.~\ref{fig:micro_swelling}b are obtained by fitting the experimental $g_2-1$, rather than its non-affine contribution. Since the experimental $g_2-1$ is dominated by the non-affine dynamics, we find essentially the same values of $\beta$ when fitting $g_2^{n.a.}-1$: $\beta = XX, XX, XX, XX, XX$, and XX for increasing $t\ge 7.2$ min}, as expected for diffusive motion~\cite{berne2000dynamic,puseyDLSchapter}.
However, for $t\ge 90$ min, both $\alpha$ and $\beta^{n.a.}$ significantly deviate from unity, approaching or even exceeding a value of $1.5$. These values are not compatible with diffusive dynamics. Rather, $\beta \approx 1.5$ has been reported for the ballistic dynamics attributed to the strain field resulting from the relaxation of internal stresses in a variety of amorphous systems~\cite{cipelletti_slow_2005,bandyopadhyay_slow_2006,bouchaud_anomalous_2008,madsen_beyond_2010}. In this case, the probability distribution of the particle displacements $\Delta r(\tau)$ at fixed $\tau$ (corresponding to fixed $\Delta \varepsilon$ in our case) is described by a Levy law, with a power-law decaying wing $\sim \Delta r^{-(\beta+1)} $ at large $\Delta r$~\cite{cipelletti2003universal}. Strictly speaking, this non-Gaussian distribution is incompatible with the assumptions underlying Eq.~\ref{eq:nonaffineg2}, which was used to calculate the mean squared displacements shown in Fig.~\ref{fig:non_affine}c. Accordingly, one should be cautious in commenting the value of the exponent $\alpha$ for the latest stages of swelling. However, it is worth pointing out that $\alpha > 1$ indicates supradiffusive behavior, thus confirming that in the late stages of swelling the microscopic dynamics are non-diffusive.

\subsection{Discussion}
\label{sec:discussion_polymer}

Although the literature on swelling gels is abundant \cite{PhysRevLett.68.2794,baselga1989effect,richbourg2020swollen,doi2009gel}, studies dealing with spherical samples are not common, especially in experiments~\cite{sun2020amino,matsumoto2004glucose,park1998thermally,barros2012surface,engelsberg2013free,bertrand2016dynamics,tanaka1979kinetics}. In this section, we briefly discuss our main findings in view of Refs.~\cite{barros2012surface,engelsberg2013free,bertrand2016dynamics,tanaka1979kinetics}, where gels of similar composition were investigated. When comparing results from different studies, it is important to keep in mind that the exact gel composition may vary across studies (or may not even be known precisely for commercially available beads), and that the experimental protocol may somehow differ (preconditioning of the gel  before the swelling experiment, nature of the solvent in which the bead is immersed, etc.). Hence, any comparison is likely to be relevant at a semi-quantitative level only.

We find that the swelling of our gel beads is, at the macroscopic level, similar to that found in previous works. The gel surface remains nearly spherical, as in Ref.~\cite{tanaka1979kinetics}, without developing the shape instability reported, e.g. in Refs.~\cite{bertrand2016dynamics,barros2012surface,engelsberg2013free}. This is consistent with the observation that shape instabilities develop in relatively rigid gels, while the elastic modulus of our beads, $G_0= 700$ Pa, is significantly smaller than that in Refs.~\cite{barros2012surface,engelsberg2013free} (27---84 kPa). The growth of bead radius follows a sub-diffusive behavior, $R\sim t^{\gamma}$ with an exponent $\gamma=0.35$, very close to $\gamma=0.38$, as predicted theoretically by Bertrand \textit{et al.}~\cite{bertrand2016dynamics}. This value is also close to, but somewhat lower than, $\gamma = 0.45$---$0.46$, as found experimentally in Refs.~\cite{barros2012surface,engelsberg2013free} for the swelling of commercial polyacrylamide hydrogel beads, and in Ref.~\cite{tanaka1979kinetics}, which measured beads of polyacrylamide gel, synthesized by using acrylamide and N-methylene-bis-acrylamide as crosslink agent.
%\MM{\cite{barros2012surface,engelsberg2013free} have a prefactor of 0.45 min$^{-\gamma}$, \cite{tanaka1979kinetics} has 0.007min$^{-\gamma}$ and we have 0.033min$^{-\gamma}$ (in all cases I have normalized the fitting parameter by the initail radius)}. 

Another similarity between the present study and previous works is the presence of a core-shell structure that rapidly forms when the gel bead is exposed to the solvent and gradually fades out as the gel tends to an asymptotic equilibrium state. In previous works, the boundary between a swollen shell and a dense core was identified by probing the local structure 
with NMR~\cite{engelsberg2013free,barros2012surface} or by shadowgraphy~\cite{bertrand2016dynamics}. Our results obtained by analyzing the spatial dependence of the scattered intensity $I$ indicate that a similar
core-shell scenario also applies to the gels studied here. We find that as the gel swells the core-shell interface propagates towards the center of the bead with an average speed of $1.1~\mu\mathrm{m}~\mathrm{s}^{-1}$, comparable to $1~\mu\mathrm{m}~\mathrm{s}^{-1}$ as measured by \cite{engelsberg2013free}. 
Similarly to Ref.~\cite{barros2012surface,engelsberg2013free}, we observe a continuous shrinking of the core. By contrast, Bertrand \textit{et al.}~\cite{bertrand2016dynamics} reported a non-monotonic evolution of the core size, which initially increased, before shrinking and eventually disappeared. This discrepancy was explained as stemming from thresholds effect for identifying the boundary, which may be different according to the experimental technique~\cite{bertrand2016dynamics}.

Interestingly, we find that tracking the position of the interface between the swollen shell and the core is not sufficient to fully account for the time evolution of the gel. Indeed, Fig.~\ref{fig:micro_swelling}b clearly shows that, as the shell propagates inwardly, the magnitude of the spatial gradient in scattering intensity corresponding to the core-shell interface becomes significantly lower. This suggest that restructuring occurs in both regions, such that the concentration difference between shell and core progressively vanishes. This scenario is in contrast with the simple picture of a static core that remains essentially unchanged until it is completely invaded by the shell, as suggested by the data shown in~\cite{engelsberg2013free,barros2012surface}. Indeed, our space-resolved data on the microscopic dynamics demonstrate unambiguously that rearrangements do occur in the core as well, from the very beginning of the swelling process (see Figs.~\ref{fig:micro_swelling}c,d). To our knowledge, this is the first experimental evidence that the core-shell scenario should not be regarded as the growth of a swollen soft layer at the expenses of an otherwise unchanged, rigid substrate. Rather, the gel evolves globally as a material in which compressive stresses propagate in the whole volume, as hypothesized by Bertrand \textit{et al.}~\cite{bertrand2016dynamics}, based on numerical results.

Finally, our PCI measurements have shed light on the dynamics associated with the microscopic rearrangements that allow for gel swelling, a quantity that was not available in previous studies. We find a striking difference between the dynamics during shrinkage (in the gel preconditioning stage) and upon swelling. While the former are essentially consistent with the homogeneous shrinkage of a perfectly elastic sphere, the latter unveil substantial non-affine microscopic rearrangements. These non-affine dynamics are reminiscent of those observed for polymer gels under shear~\cite{basu2011nonaffine} and for many other jammed systems upon applying an external strain or stress, e.g. colloidal gels and glasses~\cite{aime2018microscopic,besseling2007three,laurati2012transient,varnik2006structural} and polymer glasses~\cite{lee2009direct}. For the gels studied here, we speculate that the difference between affine dynamics upon shrinkage and non-affine dynamics during swelling may be due to the additional configurations that become accessible to the polymer chains as the gel swells, due to the increase of free volume. Interestingly, 
the nature of the non-affine mean squared displacement in the center of the bead evolves with time, from diffusive-like behavior at the onset of swelling, to supradiffusive, non-Gaussian behavior at later stages. These differences highlight, once again, the notion that the gel core should not be considered as a static, unchanged medium. They furthermore show that the microscopic dynamics exhibit a rich behavior, in spite of the smooth, subdiffusive-like evolution of the macroscopic bead size.

\section{Conclusions}
\label{sec:conclusions}

We have presented a new dynamic light scattering apparatus specifically designed for probing soft matter systems within a millimetric drop. By combining Photon Correlation Imaging~\cite{duri2009resolving}, the use of a thin incident beam, and a carefully designed collection optics, we have addressed the challenges posed by the small radius of curvature of the sample-surrounding medium interface, which acts as an additional refractive element. The setup allows for measuring the dynamics coarse-grained over regions of interest that typically correspond to sections of the incident beam of length 250-500 $\mu\mathrm{m}$. For the central ROI, the scattering angle is 90 deg; the scattering angle associated to the other ROIs depend on the refractive index contrast between the sample and the surrounding medium; typical deviations from 90 deg are up to $\pm 30$ deg. Thus, the length scales probed by the setup are of the order of a few tens of nm. Given the sample geometry, spatial resolution requires the scattered light to be collected from the side: in practice, this limits the range of accessible $q$ vectors, since it would be impossible to significantly extend the range of $q$, e.g. to smaller angles, while retaining PCI capabilities. Another limitation concerns the accessible time scales, which are dictated by the use of a CMOS detector: the smallest delay time is of the order of $10^{-4}$ s, larger than in conventional DLS, for which delay times down to a few tens of ns are available. On the other hand, the setup is sufficiently stable to probe relaxation process as slow as several thousands of seconds, which are unaccessible to conventional DLS. Finally, the multispeckle method used to calculate $g_2-1$ drastically reduces the averaging time, such that time-varying dynamics, e.g. during gelation, drying or swelling, can be thoroughly characterized. 

Measurements on a drop of non-evaporating, diluted Brownian suspension, for which spatially homogenous dynamics are expected, have shown instead a variation of the DLS relaxation time according to position within the drop. This effect, however, has been fully accounted for by considering the $x_S$ dependence of the scattering vector: we have shown that data for the suspension drop are in excellent quantitative agreement with measurements on the same sample but using a conventional cuvette and a commercial DLS apparatus. 

After validating the setup on the Brownian system, we have demonstrated some of its capabilities by studying the swelling of a polymer gel upon re-hydration. We have found an asymmetry between shrinking and swelling. Thanks to the comparison with a theoretical and numerical analysis of DLS from a homogeneous elastic sphere under strain, we have shown that during shrinking the dynamics are affine. Quite remarkably, this implies that the same deformation field holds on length scales ranging from the macroscopic drop radius down to a few tens of nm. By contrast, during swelling the microscopic dynamics are up to one order of magnitude faster than those inferred from the macroscopic affine strain field, presumably reflecting the extra degree of freedom available to the chains as their local configurations become less constrained during swelling. As in previous works~\cite{barros2012surface,bertrand2016dynamics,engelsberg2013free}, we have found that swelling is spatially heterogeneous, starting from the gel periphery and propagating inwardly as time proceeds. Our space-resolved measurements of the dynamics, unavailable in previous works, point however to a scenario more complex than expected: at the microscopic level, rearrangements and chain displacements are not limited to the swelling shell, but also occur in the core, albeit at a slower rate.  

Overall, the results presented here demonstrate the potentiality of the setup for investigating microscopic dynamics in millimeter-sized spherical samples. A key feature is the possibility of collecting data with both temporal and spatial resolution, making the apparatus an ideal tool for understanding the rich phenomenology occurring in a wide range of applications and phenomena, from the drying of drops of colloidal or biological suspensions~\cite{tsapis2005onset,milani2023double,basu2016towards,huynh2022evidence}, aggregation phenomena e.g. in the production of supraparticles for industrial applications~\cite{liu2019tuning,wooh_synthesis_2015}, to polymer drops degradation~\cite{bajpai2006investigation} or gel syneresis~\cite{wu2022effect} and swelling.

%Many other processes could benefit from this new light scattering setup. 

%%% LUCA DONE FROM HERE TO THE END

\begin{acknowledgments}
We thank J. Barbat for help in designing and building the sample environment, C. Kindelberger for help in measuring the setup stability, and C. Ligoure and P. Olmsted for enlightening discussions. We acknowledge financial support from the French Agence Nationale de la Recherche (ANR) (Grant No. ANR-19-CE06-0030-02, BOGUS). LC gratefully acknowledges support from the Institut Universitaire de France. 
\end{acknowledgments}

%%%%%%%%%%%%%%%%%%%%%%%%%%%%%%%%%%%%%%%%%%%%%
\appendix
%%%%%%%%%%%%%%%%%%%%%%%%%%
%%%%%%% APPENDIX A scattering vector

\section{Calculation of the scattering vector}\label{app:qvector}
With reference to Fig.~\ref{fig:scattering_for_theory}b, we calculate the scattering vector $q$ for a ray emerging from the drop in $P = (x_P,y_P,z_P)$ and propagating along the $y$ axis. With no loss of generality, we shall assume that the camera is placed in the $y>0$ half-space and that the scattering angle $\theta_s$, the angle of incidence $\theta_i$ and the angle of refraction $\theta_r$ are taken to be non-negative, implying $0 \le \theta_s \le \pi$, $0 \le \theta_i \le \pi/2$, $0 \le \theta_r \le \pi/2$. The unit vector associated with $P$ is 
\begin{equation}
\label{eq:P}
\hat{r}_P = \frac{x_P\hat{e}_x +y_P\hat{e}_y+z_P\hat{e}_z}{\sqrt{x_P^2+y_P^2+z_P^2}} \,,
\end{equation}
which also defines the outgoing normal to the drop surface in $P$. The angle of refraction satisfies
\begin{equation}
\label{eq:cos_angle_refraction}
\cos\theta_r = \hat{e}_y \cdot \hat{r}_P \equiv A \,.
\end{equation}
The angles of refraction and of incidence are thus given by
\begin{equation}
\label{eq:angle_refraction}
\theta_r = \arccos A \,,
\end{equation}
\begin{equation}
\label{eq:angle_incidence}
\theta_i = \arcsin \left( \frac{n_m}{n} \sin\theta_r\right )\,,
\end{equation}
where in writing Eq.~\ref{eq:angle_incidence} we have used Snell's law, with $n$ and $n_m$ the refractive indices of the drop and of the surrounding medium, respectively.

For the special case $n=n_m$, the scattering angle is $\theta_s = \pi/2$ for all $x_P$. Similarly, $\theta_s = \pi/2$ for $x_P=0$ and arbitrary $n$, $n_m$. For these cases, $\hat{k}_o = \hat{e}_y$, with $\hat{k}_o$ the unit vector along the direction of the scattered light in the drop. It is then straightforward to calculate $\mathbf{q}$ using Eq.~\ref{eq:qvector}. For all the other cases, it is convenient to express $\hat{k}_o$ as a linear combination of the two unit vectors $\hat{r}_P$ and $\hat{e}_y$:
\begin{equation}
\label{eq:ko_linear_combination}
\hat{k}_o = a \hat{r}_P + b\hat{e}_y\,,
\end{equation}
with $a$ and $b$ two coefficients to be determined. Figure~\ref{fig:scattering_for_theory}b shows that
\begin{equation}
\begin{aligned} 
\label{eq:ko_linear}
\hat{k}_o \cdot \hat{r}_P &=  \cos\theta_i \\ 
\hat{k}_o \cdot \hat{e}_y &=  \cos(\theta_r - \theta_i) \,.
\end{aligned}
\end{equation}
By inserting Eqs.~\ref{eq:cos_angle_refraction}~and~\ref{eq:ko_linear_combination} into Eqs.~\ref{eq:ko_linear}, one has
\begin{equation}
\begin{aligned} 
\label{eq:a_b_system}
a + bA  &=  \cos\theta_i \\ 
aA + b &=  \cos(\theta_r - \theta_i) \,.
\end{aligned}
\end{equation}
For the ease of writing, we define $B = \cos\theta_i$ and $C = \cos(\theta_r - \theta_i)$ and solve Eqs.~\ref{eq:a_b_system} for $a$ and $b$, finding
\begin{equation}
\begin{aligned} 
\label{eq:a_b}
a &= \frac{B-AC}{1-A^2} \\
b &= \frac{C-AB}{1-A^2} \,.
\end{aligned}
\end{equation}
Equations~\ref{eq:a_b}, together with $\textbf{k}_o = 2\pi n \lambda^{-1} \hat{k}_o$ and Eqs.~\ref{eq:P} and~\ref{eq:ko_linear_combination} allow for calculating the scattering vector for scattered light being imaged on the camera detector after emerging from the drop in $x_P$ and having propagated along the $y$ direction, for the general case $\hat{k}_o \ne \hat{e}_y$ .

%%%%% APPENDIX B: analytical g2-1 for affine motion

\section{Determination of analytical $g_2-1$}\label{app:analyticalg2detal}
%\LR{difficult to follow for me, but it is an appendix ...}

We aim at calculating an analytical expression of the intensity correlation function $g_2-1$ for the light scattered by a contracting or expanding sphere and emerging in point $P$, as shown in Fig.~\ref{fig:scattering_for_theory}. We assume in the following that the strain field within the sphere is that of an ideal solid, i.e. we calculate only the contribution to $g_2-1$ due to affine displacements. As shown in Fig.~\ref{fig:scattering_for_theory}, light emerging in $P$ is scattered by all points on the $PT$ segment. In our experiments, the incident beam is Gaussian, with a thin cross section (beam waist $w_0 \approx 60~\mu$m), much smaller than the drop radius $R$. Thus, most of the scattered light originates from a small portion of $PT$ centered around point S on the $x$ axis. However, to ease the calculation, it is convenient to consider the whole segment $PT$. Since $w_0 << R$, refraction effects on the incoming beam may be neglected, since they are only relevant for the portions of the incoming beam far from the $x$ axis, whose intensity is vanishingly small. We thus assume that the incident beam within the drop propagates along the $\hat{e}_x$ direction, regardless of the distance from the $x$ axis. This insures that light scattered by any point along $PT$ has the same wave vector $\textbf{k}_o$ and thus the same scattering vector $\textbf{q}$.

We start by determining the coordinates of $P$ and $T$. Experimentally, $g_2-1$ is calculated for all pixels belonging to a ROI, i.e. for a set of pixels identified by $x_P$ and $z_P$, the coordinates of $P$ in the $(x,z)$ plane, which are obtained directly from the CMOS image. Since $P$ lies on the surface of the drop, one has 
\begin{equation}
\label{app:Pcoordinates}
\textbf{OP} = x_P \hat{e}_x + \sqrt{R^2-(x_P^2 + z_P)^2}  \hat{e}_y + z_P \hat{e}_z \,.
\end{equation}
The coordinates of point $T$ are found by imposing that $PT$ is colinear with $\hat{k}_o$ and that $T$ lies on the surface of the sphere, in the $y<0$ half-space (see Fig.~\ref{fig:scattering_for_theory}). One finds
\begin{equation}
\label{app:Tcoordinates}
\textbf{OT} = \textbf{OP} - \left(2 \textbf{OP} \cdot  \hat{k}_o \right) \hat{k}_o \,.
\end{equation}
Equation~\ref{app:Tcoordinates}, together with Eq.~\ref{app:Pcoordinates} and the expression of $\hat{k}_o$ provided in Appendix~\ref{app:qvector} fully determines the coordinates of $T$ for any experimentally chosen $(x_P,z_P)$.

To find an analytical expression for $g_2-1$, we need to write the scattered intensity $I$ originating from an arbitrary point $\mathbf{x}$ along $PT$ as a function of the parameter $l$, see Eq.~\ref{eq:intesity0}. Using Eqs.~\ref{eq:pointinthesphere}-\ref{eq:intesity0} and the coordinates of $P$ and  $T$ given above, one finds
\begin{equation}
   I(\mathbf{x}) \propto \exp\biggl\{ - ( D l^2 + E l +F  )/(w_0^2/2)     \biggr\}
    \label{eq:intesityappendix}
\end{equation}
where 
\begin{equation}
\begin{aligned} 
\label{eq:constants_Gaussian}
D &= (y_T-y_P)^2 + (z_T-z_P)^2 \\
E &= 2z_T(y_T-y_P) + 2y_T(z_T-z_P) \\
F &= y_T^2 + z_T^2
\,,
\end{aligned}
\end{equation}
and where we have omitted all unnecessary prefactors. Finally, we need an expression for the displacement field $\Delta\textbf{u}(\textbf{x},\Delta\varepsilon)$ in $\mathbf{x}$ corresponding to a change of macroscopic strain $\Delta\varepsilon$. We use a simplified version of Eq.~\ref{eq:displacementfield}, where we drop the second term in the r.h.s., which is typically negligible, finding
\begin{equation}
\Delta\textbf{u}[\textbf{x}(l),\Delta\varepsilon] = l \Delta\varepsilon (\textbf{OT}-\textbf{OP})   +\Delta\varepsilon \textbf{OT}  \,.
\label{eq:Deltau_simplified}
\end{equation}
The projection of the displacement onto the scattering vector is then
\begin{equation}
\Delta\textbf{u}\cdot\textbf{q}= l \Delta\varepsilon \left[ \textbf{q}\cdot(\textbf{OT}-\textbf{OP})   \right ]+\Delta\varepsilon \big( \textbf{q}\cdot\textbf{OT}     \big)=
\tilde{q} l + K
\label{eq:phasesimplified}
\end{equation}
where we have introduced the strain-dependent generalized wave vector $\tilde{q}=\Delta\varepsilon  \left[\textbf{q}\cdot(\textbf{OT}-\textbf{OP)}   \right]$ and defined $K=\Delta\varepsilon \big( \textbf{q}\cdot\textbf{OT}\big)$. By inserting Eqs.~\ref{eq:intesityappendix} and~\ref{eq:phasesimplified} into Eq.~\ref{eq:analiticalg2_2} and omitting temporarily any normalization factor, we obtain:
\begin{equation}
g_2-1 \propto \left| \int_{0}^{1} e^{-i \tilde{q} l} \exp \left\{ - ( D l^2 + E l +F  )/(w_0^2/2)     \right\} \mathrm{d}l \right|^2
\end{equation}
which, by extending to $\pm\infty$ the integration limits with no significant change since the integrand is vanishingly small out of $[0,1]$, is the Fourier transform of a generalized Gaussian function. This Fourier transform has an analytical solution~\cite{bateman_tables_1954}:  
\begin{equation}
\label{eq:analytical}
    g_2-1 = e^{-2\frac{(w_0^2/2)\tilde{q}}{4D}} \,,
\end{equation}
where we have used the correct normalization. Note that in the above expression $g_2-1$ depends on $\Delta \varepsilon$ through $\tilde{q}$. Over a time laps short enough for the strain rate $\dot{\varepsilon}$ to be considered constant, one has $\Delta \varepsilon = \tau \dot{\varepsilon}$. Under these conditions, Eq.~\ref{eq:analytical} is of the form of Eq.~\ref{eq:g2brownianstrect} of the main text with $\beta=2$ and
\begin{equation}
\label{eq:tauE}
    \tau_E = \frac{2\sqrt{2} \sqrt{(y_T-y_P)^2+(z_T-z_P)^2}}{\dot{\varepsilon} \left[\textbf{q}\cdot(\textbf{OT}-\textbf{OP})\right]w_0} \,,
\end{equation}
\begin{equation}
\label{eq:tauD}
    \tau_D = \frac{\tau_E}{2\sqrt{2}}\Gamma \left({\frac{1}{2}}\right)  \approx 0.6267 \tau_E \,.
\end{equation}
\section{Determination of the core-shell boundary in a bead of polymer gel}\label{app:Igel}

%%%%%%%%% FIGURE ANALYSIS CORE-SHELL %%%%%%%%%%%%%%%
\begin{figure}
\includegraphics[width=1\linewidth]{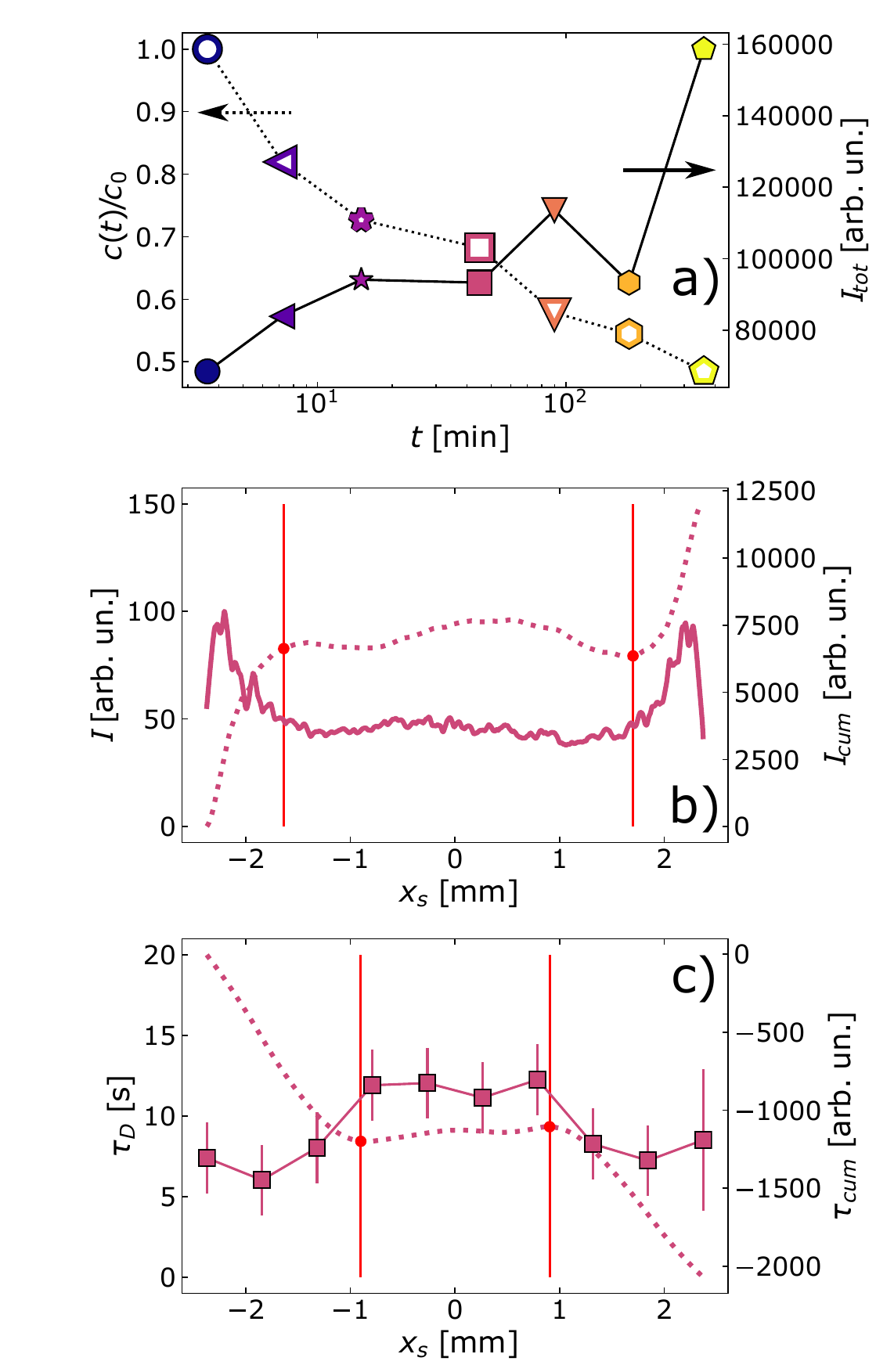}% Here is how to import EPS art
\caption{\label{fig:Igel} (a): Relative change of the average monomer concentration upon swelling of the gel of Sec.~\ref{sec:polymergel} (symbols and left axis), together with the time evolution of the scattered intensity integrated over the whole scattering volume (solid symbols and right axis). (b) Scattered intensity profile (left  axis and solid line) and its integral up to position $x_S$ (right axis and dotted line). (c) Relaxation time as a function of position (left axis and symbols), together with its integral up to position $x_S$ (right axis and line). In b) and c), $t=45$ min and the vertical lines and red dots indicate the core-shell boundaries.}
\end{figure}
%%%%%% END FIGURE %%%%%%%%%%%%%%%%%%%%%%%%%%%%%%%%%%

We first show that, for our system, an increase of the scattered intensity $I$ corresponds to a decrease in monomer concentration, e.g. to gel swelling. This is demonstrated by Fig.~\ref{fig:Igel}a, where we plot on the left axis the relative change of the average monomer concentration due to swelling, $c(t)/c_0 = R_0^3/R(t)^3$, where the subscript 0 indicates quantities measured at time $t=0$, and on the right axis the time dependence of $I_{tot}$, the scattered intensity integrated over the whole scattering volume. The two quantities follow an opposite trend.

In order to determine the core-shell boundary, we apply a similar method to analyze  the $x_S$ dependence of the static scattered intensity $I$ and of the relaxation time $\tau_D$, respectively. For each of the $I(x_S,t)$ curves shown in Fig.~\ref{fig:micro_swelling}b, we plot the $x_S$ dependence of $I_{cum}(t) = \int_{x_{min}}^{x_S} [I(x',t)-I_C]\mathrm{d}x'$, where $I_C$ is the intensity value in the plateau region around the bead center where the scattered intensity is roughly independent of position. Figure~\ref{fig:Igel}b shows a typical $I_{cum}$ curve, for $t=45$ min (dotted line and right axis), together with the corresponding $I(x_S)$ data (solid line and left axis).  $I_{cum}$ initially raises steeply, in the $x_S<0$ part of the shell, it then levels off in the core and raises again in the $x_S>0$ part of the shell. By inspecting the profile of $I_{cum}$, we determine the core-shell boundaries, shown by the red dots and vertical lines in Fig.~\ref{fig:Igel}b. From their values, we determine the average position of the interface and its uncertainty, reported in Fig.~\ref{fig:micro_swelling}b.

We use the same approach for analyzing the $\tau_D(x_S)$ data of Fig.~\ref{fig:micro_swelling}d:  
Figure~\ref{fig:Igel}c shows $\tau_{cum}(t) = \int_{x_{min}}^{x_S} [\tau_D(x',t)-\tau_C]\mathrm{d}x'$ (right axis), where $\tau_C$ is the relaxation time in the plateau region around the bead center \footnote{We use linear interpolation to estimate $\tau_D$ in between the available data points}. The corresponding $\tau_D(x_S)$ data are shown as symbols (left axis) and the red dots and vertical lines mark the core-shell interfaces.

\nocite{*}

\bibliography{tech_paper}% Produces the bibliography via BibTeX.

\end{document}

\preprint{APS/123-QED}

\title{Drying of a colloidal drop - Supplemental Information}

\author{Matteo Milani}
% \altaffiliation[Also at ]{}%Lines break automatically or can be forced with \\
\author{Ty Phou}%
\author{Christian Ligoure}%
\author{Luca Cipelletti}%
\email{luca.cipelletti@umontpellier.fr}
\author{Laurence Ramos}%
 \email{laurence.ramos@umontpellier.fr}

\affiliation{%
 Laboratoire Charles Coulomb (L2C), Universit\'e Montpellier, CNRS, Montpellier, France\\
}%

\maketitle

\section{Methods}

\subsection{Imaging}

For the imaging of Fig.3 of the main text we use the same optical set-up the we use for light scattering adding a white light source on the back of the drop. The combination of the vertical slit and of the diaphragm reduces the numerical aperture of the system to $10^{-2}$.

\subsection{Light scattering set-up}
\subsubsection{Optics}

The first lens ($L_1$ in Fig.1a of the main paper) has a diameter $65$ mm and focal length $64$ mm.
The second lens ($L_2$ in Fig.1a of the main paper has a diameter of $40$ mm and focal length $82$ mm. The vertical slit has an aperture of $4$ mm and the diaphragm has a diameter of $12$ mm. The combination of the slit and diaphragm selects light with angular intervals $\Delta\theta_{xy} =1.1$ deg, respectively  $\Delta\varphi_{yz} =3.3$ deg, in the $(x,y)$, respectively $(y,z)$, plan. 

\subsubsection{Measurement of the beam waist}

To determine the beam waist $w$ of the laser light illuminating the sample, we replace the drop with a frosted glass placed perpendicular to the incoming beam, at $x=0$, i.e. in the position occupied by the center of the drop during the drying experiments. We record the speckle pattern generated by the light scattered by the frosted glass using a CMOS camera placed in the focal plane of a converging lens with focus $f=4$ cm along the $x$ axis. We calculate the spatial autocorrelation of the intensity pattern, $ C_I( x,y)= \langle I(x_1,y_1)I(x_2,y_2)\rangle$. The autocorrelation function can be written as ~\cite{goodman2007speckle}
\begin{equation}
    C_I(x,y)=\rm{exp} \bigg[- \frac{\pi^2  w^2}{2f^2\lambda^2}(x^2+y^2)\bigg]
\end{equation}

The beam waist size $w=59.5\pm0.1$ $\mu m$ is extracted from the fit of the spatial autocorrelation with a Gaussian function~\cite{goodman2007speckle}.

\subsubsection{Camera features}
The light scattered by the sample is collected either on one detector or on two detectors (Camera 1 and Camera 2 in Fig.~1a in the main manuscript). 
Camera 1, which is used in all experiments, is a Basler camera acA2000 340km, with a detector size of $2048$ pixels $\times 1088$ pixels and a pixel size of $5.5$ $\mu$m. We typically reduce the resolution of Camera 1 down to $2048$ pixels $\times 512$ pixels. With this camera it is possible to collect data over a very long period of time, from hours or days, but the acquisition is limited to a maximum rate of $200$ fps, thus preventing any fast dynamics to be probed. To overcome this limitation, we use a beam splitter to reflect part of the scattered light to a second detector (Camera 2). For Camera 2, we use a Phantom Miro M310 with a detector $1280 \times 800$ pixel$^2$ (pixel size  $20$ $\mu$m). We usually reduce the resolution of this camera to $896$ pixels $\times 184$ pixels, such that the maximum acquisition rate is $19000$ fps.

\subsubsection{Propagation vector $\overline{k_o}$ }
To get precisely the scattering angle $\theta_{s}$ we need to map the position of a light spot on the camera with its position in the sample. As show in Fig.~\ref{fig.scatt_vs_q}, a point on the camera with coordinates $P_c = (x,y_c,-z)$ has travelled in the direction $\widehat{a}$ from the point on the surface of the sphere $P_s=(x,\sqrt{R^2-x^2-z^2},z)$, where $R$ is the radius of the sphere that has center in $\overline{0}$. Due to the index refraction mismatch, the ray that propagates from $P_s$ has been deflected at the sample-medium interface following the Snell law:
\begin{equation} 
\label{eq:refraction}
\begin{split}
\theta_2& = \arccos(\widehat{a}\cdot\widehat{r}) \\
\theta_1& = \arcsin\big(\frac{n_2}{n_1}\sin(\theta_2)\big)\\ 
\end{split}
\end{equation}
the propagation vector $\overline{k_o}$ can be obtained by geometrical relationship $\overline{k_o}=\alpha \widehat{a} + \beta \widehat{r}$, where:

\begin{equation} 
\label{eq:refraction}
\begin{split}
\alpha& = \cos( \theta_2-\theta_1 ) - \beta  \cos( \theta_2 ) \\
\beta& = \frac{ \cos( \theta_1 ) -  \cos( \theta_2-\theta_1 ) \cos( \theta_2 ) }{  ( 1 - \cos( \theta_2 )^2 )}\\ 
\end{split}
\end{equation}

\begin{figure}
\centering
\includegraphics[width=0.8\linewidth]{FIGURES_SI/3d_with_angles.pdf}
\caption{3D construction of a water based sample $n_1=1.33$ disperse in air $n_2=1$. The round black dot represent the scatterer the deflects the light coming from the laser in the direction $\widehat{k_o}$. The light arrives at the interface in the point $P_s$ and is deflected according to the Snell law in the direction $\widehat{a}$ until reaching the camera in the point $P_c$. The green cylinder is the laser.}
\label{fig.scatt_vs_q}
\end{figure}

%For the drop in air the decay time $\tau_D$ increases with $x$, while $\tau_D$ decreases with $x$ for the sample surrounded by oil. 

\subsubsection{Data acquisition and laser alignment}

In order not to have the laser deflected by the sample's interface two conditions are required: the beam waist $w$ should be much smaller than the characteristic size of the sample and the laser beam should hit the surface of the drop at $90^{\circ}$ with respect to a tangent plane of the spherical sample. 

From an experimental point of view, the drop sits on the surface on the top of a stage that can translate in the $z$ direction. In this a way, the position of the sample can be adjusted so that the laser beam hits the spherical sample in its equatorial plane. Since during the drying the drop evaporates, the sample is translated upwards in order to maintain the condition of perpendicularity. The timing with which this operation is done depends on the evaporation rate. It vary from $15$ min for $Pe=57$ to $60$ min for $Pe=9$. Each time, the drop is translated from about $100$ $\mu$m.  

\subsection{Sample environment}

%To prepare the suspension at 10mM with a volume fraction $\phi = 0.3$ we need to concentrate the Ludox of the mother batch at a concentration such as is possible to add a KCl solution. The 10mM of salt concentration are calculated respect to the volume of the solvent before drying, excluding the the volume of silica particles. Due to evaporation the salt concentration is expected to reach 15mM for the sample dried at Pe= 36 and 17mM for the sample dried at Pe=13 .We check that with these salt concentrations the colloids do not aggregate. 

\subsubsection{Preparation of the hydrophobic surface}
To produce the hydrophobic surface onto which the drop sits, a water ethanol solution of Trimethoxy(octadecyl)silane ($20$ mg/ml) is deposited on a glass slide and heated in an oven ($1$ h ramp from $30^{\circ}$C to $70^{\circ}$C, and then $12$ h at $70^{\circ}$C). The contact angle of the drop on the surface, measured by imageJ, is around $120$ deg. 

\subsubsection{Humidity control}

Table $1$ list the compounds used to control the relative humidity (RH) of the chamber that contains the sample, the characteristic evaporation time ($\tau_{\rm{ev}}$) and the related Peclet numbers ($Pe$). RH is measured with the humidity sensors (SENISOR SEK-SensorBridge) and $Pe$ is evaluated as indicated in the main manuscript.

%%%%%%%%%%%%%%%%  TABLE 1 %%%%%%%%%%%%%%%%%%%%
\begin{center}
\begin{table}[h!]
\begin{tabular}{||c c c c ||} 
 \hline
   & RH [$\%$] & Pe & $\tau_{\rm{ev}}$ [min]\\ [0.5ex] 
 \hline\hline
 H$_2$O & 94 & 9 & 732\\
  \hline
 KCl & 85 & 10  & 694\\ 
 \hline
 NaCl & 75 & 13  & 476\\
 \hline
 K$_2$CO$_3$ & 43 & 25 & 248\\
 \hline
 MgCl$_2$ & 33 & 36  & 186\\
 \hline
 A.S. & 20 & 57 & 112\\
 
 \hline
\end{tabular}
\caption{List of chemicals used to controlled the relative humidity, the corresponding time for evaporation, $\tau_{\rm{ev}}$, and Peclet number, $Pe$. The salts are used in saturated solutions. A.S. is a commercial amorphous silica named also as silica xerogel. }
\label{table:1}
\end{table}
\end{center}

\bibliography{apssamp}